\documentstyle[emulateapj,psfig]{article}
\newcommand{\be}{\begin{equation}}
\newcommand{\ba}{\begin{eqnarray}}
\newcommand{\ee}{\end{equation}}
\newcommand{\ea}{\end{eqnarray}}  
\newcommand{\etal}{et al.\ }
\def\E55{E_{55}}  
\def\bstrip{b_{\rm strip}}  
\def\bcool{b_{\rm cool}}  
\def\tcool{t_{\rm cool}}  
  
\def\gtsima{$\; \buildrel > \over \sim \;$}
\def\ltsima{$\; \buildrel < \over \sim \;$}
\def\gsim{\lower.5ex\hbox{\gtsima}}
\def\lsim{\lower.5ex\hbox{\ltsima}}
\def\simgt{\lower.5ex\hbox{\gtsima}}
\def\simlt{\lower.5ex\hbox{\ltsima}}
\def\simpr{\lower.5ex\hbox{\prosima}}

\def\CII{C{\sc ~ii}}
\def\FeII{Fe{\sc ~ii}}
\def\SiII{Si{\sc ~ii}}
\def\msun{{M_\odot}}
\def\kms{\,{\rm km\,s^{-1}}}

\def\eg{{\frenchspacing\it e.g. }}

\begin{document}
\title{Triggering the Formation of Halo Globular Clusters with Galaxy Outflows}

\author{Evan Scannapieco\altaffilmark{1}, Jon Weisheit\altaffilmark{2},
\& Francis Harlow\altaffilmark{3}}
\altaffiltext{1}{ Kavli Institute for Theoretical Physics, Kohn Hall,
UC Santa Barbara, Santa Barbara, CA 93106}
\altaffiltext{2}{ Theoretical Division, Plasma Theory Group T-15,
K717, Los Alamos National Laboratory, Los Alamos, NM 87545}
\altaffiltext{3}{ Theoretical Division, Fluid Dynamics Group T-3,
B216, Los Alamos National Laboratory, Los Alamos, NM 87545}

\begin{abstract}

We investigate the interactions of high-redshift galaxy outflows with
low-mass virialized clouds of primordial composition.  While atomic
cooling allows star formation  in objects with virial temperatures
above $10^4$K,  ``minihalos'' with virial temperatures below this
threshold are generally unable to form stars by themselves.  However,
the large population of high-redshift starburst galaxies may have induced
widespread star formation in neighboring minihalos, 
via shocks that caused intense cooling both through
nonequilibrium $H_2$ formation and metal-line emission.  Using a
simple analytic model, we show that the resulting star clusters
naturally reproduce three key features of the observed population of
halo globular clusters (GCs).  First, the $10^4$K maximum virial
temperature directly corresponds to the $\sim 10^6 \msun$ upper limit
on the stellar mass of such clusters, a feature that can not be
explained by any GC destruction mechanism.  Secondly, the momentum
imparted in such interactions is sufficient to strip the gas from its
associated dark matter halo, explaining why GCs do not reside in the
dark matter potential wells that are ubiquitous in galaxies.  Finally,
the mixing  of ejected metals into the primordial gas provides a
straightforward mechanism to explain the  $\sim -0.1$ dex homogeneity
of  stellar metallicities within a given GC, while at the same time
allowing for a large spread in metallicity between different clusters.
To study the possibility of such ``fine grained'' mixing in detail,
we use a simple one-dimensional  numerical model of turbulence
transport to simulate mixing in cloud-outflow interactions. We find
that as the shock shears across the side of the cloud, Kelvin-Helmholtz
instabilities arise, which cause turbulent mixing of enriched material
into $\gtrsim 20\%$ of the cloud.  Such estimates ignore the likely
presence of large-scale vortices, however, which would further enhance
turbulence generation.  Thus the global nature of mixing in these
interactions is multidimensional, and quantitative predictions must 
await more detailed numerical studies.

\end{abstract}

%\keywords{intergalactic medium -- galaxies: interactions }

\section{Introduction}

Most of the Milky Way's $\sim$ 150 globular star clusters are old.
Their ages have been reliably determined from main sequence turnoff
points, with the firm result that the majority formed  10-13 Gyr ago
and thus are coeval with the oldest stars in the Galaxy (Krauss
\& Chaboyer 2003, and references cited therein).   Furthermore, the
existence of similar systems in many other galaxies suggests that
these objects represent an essential aspect of the epoch of galaxy
formation.  Seminal papers by Peebles \& Dicke (1968), Searle \& Zinn
(1978), and Fall \& Rees (1985) spurred numerous lines of
investigation into this relationship and produced scenarios in which
globular clusters (hereafter GCs) formed either before, during, or
shortly after the development of the galaxies that now host them.
Unfortunately, improved theory and simulations, and especially, new
observations have made most of the earlier scenarios untenable, or at
least unlikely.  Hence, the question of what triggers the formation of
globular clusters has yet to be satisfactorily answered.

One key development for this subject was the work of Zinn (1985), who
established that two distinct populations of Galactic GCs exist:  (1)
relatively metal-rich clusters, with [Fe/H] $>$ -0.8, which share
spatial and kinematic properties with the Galaxy's thick disk or
bulge; and (2) relatively metal-poor clusters, with -2.5 $<$ [Fe/H]
$<$ -0.8, which appear to be  part of the halo.  Further work has
indicated a similar dichotomy in other galaxies (\eg Ashman \& Bird
1993;  Forbes, Brodie, \& Huchra 1997; Beasley \etal 2000;
Larsen \etal 2001),  which
suggests two different modes of GC formation.   In the metal-rich
case, formation is likely to be an ongoing process.   External
galaxies, often in the midst of strong tidal  interactions,  have been
observed to host OB associations with various properties expected of
young GCs -- they are compact, bright, bluish, and are estimated to
have masses $> 10^4 M_\odot$ (\eg Whitmore \& Schweizer 1995;
Schweizer \etal 1996).    Furthermore, it has recently been realized
that galaxies  can add to their system of globulars by stripping GCs
from dwarf satellite galaxies, as seems to be occurring now in
connection with the Milky Way's Saggitarius dwarf (Ibata \etal 2001;
Yoon \& Lee 2002).  In low-metallicity GCs, only ages $\ge 10$ Gyr are
seen, and characteristics appear to be more uniform across galaxy
types and sizes (see however, Strader, Brodie, \& Forbes 2004).  While
the reader is referred to  the recent monograph by Ashman \& Zepf
(1998) for a more detailed comparison between these populations, our
focus in this work will be to develop a formation model only for halo
globular clusters.

There are three essential features of these objects that any formation
scenario  must explain or accommodate.  The first of these is the
remarkable chemical homogeneity of iron-peak elements exhibited by
stars within a given cluster: typically, the dispersion in [Fe/H]
is less than 0.1 dex (\eg Suntzeff 1993).  This is particularly
extraordinary in the case of the oldest clusters, in which the
star-forming gas must have attained its metallicity before the universe
was a billion years old.  This condition divides current formation
theories for halo GCs into `pre-enrichment' and `self-enrichment'
types.  In the former scenario (\eg Elmegreen \& Efremov 1997; Bromm
\& Clarke 2002), GCs  were formed out of gas that had already been
homogeneously enriched by a previous generation of supernovae (SNe).
Here the key question  is just exactly what that population was,  why
it played only a secondary role in the formation history of the GC,
and how it could have enriched this material on very short time
scales.  In the self-enrichment picture, on the other hand, the
protocluster cloud was enriched by one or more supernova events
occurring within it  (\eg Brown, Burkert, \&  Truran 1995; Cen 2001;
Nakasato, Mori, \& Nomoto 2000; Beasley \etal 2003; Li \& Burstein 2003).  
In this case, the key problems are that this self-enrichment must occur
extremely homogeneously and that the kinetic energy corresponding to
these SNe can be enough to unbind the gaseous proto-cluster (Peng \&
Weisheit 1991).  Note however that the distribution of metallicities
{\em among different} halo globular clusters is quite large, and
approximates a Gaussian with a mean of [Fe/H] $\sim
-1.59$ and a dispersion of [Fe/H] $\sim 0.34$ (Ashman \& Zepf 1998).

A second constraint comes from observations of the tails of stars that
are in the process of being stripped from GCs by the tidal field of
the Galaxy (Irwin \& Hatzidimitriou 1993; Grillmair \etal 1995).  If
globular clusters contained substantial halos of dark matter, this
increased gravitational potential would have a large impact on these
tidal losses.  In fact, no evidence of dark matter halo suppression is
seen, placing a strong upper limit of 2.5  on the ratio of total mass
to  stellar mass of these objects (Moore 1996; see however
Maschenko \& Sills 2004).  This is in direct
contrast with galaxies, which exhibit total mass to stellar mass
ratios $\sim 100$ (\eg Marinoni \& Hudson 2002).   Thus the mechanism
that formed GCs is  likely to be fundamentally different from the
dark-matter driven collapse that is believed to govern the formation
of galaxies, as it  is much easier to separate dark matter from gas
than from  point-like bodies such as stars (see, however, Bromm \&
Clarke 2002).

The final constraint on GC formation is related to their
mass distribution, which is well described as
a Gaussian in $\log_{10}(M_\star/\msun)$ with a mean 
$\left< M_\star \right> \sim 10^5 \msun$ and 
a dispersion of $0.5$ (\eg Armandroff 1989).  This issue is most lucidly
illustrated by the classic ``survival triangle'' in the mass-radius 
plane of globular clusters, which is bounded by the long-term
disruption processes that act on Galactic GCs (Fall \& Rees 1985). 
Here the minimum radius as a function of mass is determined by mechanical 
evaporation (\eg Spitzer and Thuan 1972) while the maximum radius
as a function of mass is bounded by shocking that
occurs when a cluster passes through the high-density Galactic disk
(\eg Ostriker, Spitzer, \& Chevalier 1972).   While both these constraints
seem to provide good agreement with the observed GC 
minimum stellar masses and  sizes, dynamical friction, which is 
the only Galactic mechanism limiting the {\em maximum} stellar mass,
only operates at masses $\geq 10^{7.5} \msun$.
Thus it appears that the $\sim 10^{6} \msun$ 
upper mass cut of globulars is {\em not} set 
by  any known destruction mechanism (\eg Gunn 1980; Caputo \& Castellani
1984), but rather represents an intrinsic 
property of the population of gaseous proto-clusters 
(Peng \& Weisheit 1991).

In the high-redshift universe, nature has provided us with just such a
population.  Because atomic line cooling is only effective at
temperatures $\geq 10^4$K, collapsed clouds of material with virial
temperatures below this threshold must radiate energy through dust and
molecular line emission.  While the levels of $H_2$ left over from
recombination are sufficient to cool gas in the earliest structures
(\eg Abel, Bryan, Norman 2002; Bromm, Coppi, \& Larson 2002) the
resulting 11.20-13.6 eV photon emission from the stars in these
objects (\eg Haiman, Rees, \& Loeb 1997; Ciardi \etal 2000) is likely
to have quickly dissociated this primordial $H_2$. Thus a generic
prediction of current structure formation models  is  a large
population of $T_{\rm vir} \leq 10^4$K, $M_{\rm gas} \leq 10^{6.5}
\msun$ virialized clouds  of gas and dark matter that are unable to
form stars until they interact with other objects.  In fact, it was
the similarity between these so-called  ``minihalos'' and the globular
cluster population that led Cen (2001) to propose that convergent
ionization fronts during cosmological reionization might be able to
transform these objects into GCs.  Yet such a scenario begs the
question of how globulars were enriched  with metals, and it is not at
all clear that the first galaxy-minihalo interactions were radiative.

Observations of high-redshift starburst galaxies have uncovered large
numbers of outflows, both in optical and infrared measurements at $3
\lesssim z \lesssim 4$ (Pettini \etal 2001) and in optical
observations of lensed galaxies at  $4 \lesssim z \lesssim 6.5$ (Frye,
Broadhurst, \& Benitez 2002; Hu \etal 2002).   In addition,
the classic picture of reionization is of a two stage process, which  
begins with
individual sources ionizating their immediate  surroundings and  ends
in a rapid ``overlap'' phase in  which neighboring H II region join
together, quickly ionizing the remaining neutral regions (\eg Gnedin
2000).    Thus it is possible that minihalos that were close to
starbursting  galaxies during the first stage of reionization could
have been impacted by outflows prior to reionization fronts, depending
on which was first able to escape from high-redshift starbursts.  
While numerical simulations have yet to yeild a definitive answer to
this question, there are suggestions that the I-fronts in $\gtrsim 10^7
\msun$ starbursts may be $D$-type at small radii and slow to expand
(Kitayama \etal 2004), and that shell material swept up by outflows is
effective at trapping ionizing radiation (Fujita \etal 2003).
And while eventually shell fragmentation allowed such ionizating 
radiation to escape,  even then, large regions may have remained 
``shadowed'' by the fragments.

It is possible, therefore, that a large number of minihalos were first
impacted by outflows.  Unlike radiation fronts, which 
typically boil away the minihalo gas (\eg Shapiro \etal
2004), shock interactions will result in intense cooling through both
nonequilibrium formation of $H_2$ (Mac Low \& Shull 1986; Shapiro \&
Kang 1987; Palla \& Zinnecker 1988; Ferrara 1998; Uehara \& Inutsuka
2002) and the mixing in of metals with ionization potentials below
13.6 eV (Dalgarno \& McCray 1972), thereby initiating the formation of
large numbers of stars.  Similar shock-induced star formation has been
observed and simulated in low-redshift intergalactic clouds impacted
by radio jets (van Breugel \etal 1985; Fragile \etal 2004).  Further,
the efficient dispersal of stellar nucleosynthetic products at high
redshift is required by observations of metals (at levels $Z \sim
10^{-4}$ to $10^{-2} Z_\odot$) in the metal-poor stars of the Galaxy's
halo (\eg Freeman \& Bland-Hawthorn 2002), in the hot gas within
galaxy clusters (\eg Renzini 1997; Peterson \etal 2003), and in the
intergalactic clouds producing quasar absorption line features (\eg
Tytler \etal 1995; Rauch \etal 1996; Songaila 2001).

Additionally, this formation trigger offers a plausible
explanation of why individual GCs do not have dark matter halos today.
Previous work (Scannapieco, Ferrara, \& Broadhurst 2000) has shown
that gas in growing pre-virialized density perturbations is 
vulnerable to the influence of outflows.
Here the dominant mechanism is ``baryonic stripping'' in which
the gas is accelerated above the escape velocity  and ejected
from the associated dark matter perturbation.  While this has the
effect of completely suppressing star formation in diffuse
pre-viralized regions, such interactions 
may strip the dense gas from virialized minihalos 
while at the same time enhancing their densities to form 
gravitationally-bound clouds.  

In this paper we carry out a two-part study to 
explore this scenario in detail. First, using a simple
analytical model, we examine heating, cooling, and momentum 
transfer in outflow-minihalo interactions, and determine the
general properties of star clusters formed by this process.
Second, we use one-dimensional numerical models to examine the
turbulence-driven mixing that happens when metal-rich supernova
ejecta encounters gas of primordial composition.  
Our work builds on recent analyses involving 
one of us (Thacker, Scannapieco,  \& Davis 2003; Scannapieco, Schneider,
\& Ferrara 2003)  in which basic outflow properties,
global enrichment patterns, and heavy-element yields from Population III
supernovae were determined.

The structure of this work is as follows.  In \S2 we outline a general
model for galaxy outflows and high-redshift minihalos, and in \S3 we
construct simple analytical estimates for the fate of the minihalo gas
subjected to an outflow as a function of model parameters.  In \S4 we
use one-dimensional numerical turbulence models to examine the mixing
of metals into the minihalo gas.  Conclusions are given in \S5.

\section{General Framework}

Driven by measurements of the Cosmic Microwave Background, the number
abundance of galaxy clusters, and high-redshift supernova distance
estimates (eg.\ Spergel et al.\ 2003; Eke \etal 1996; Perlmutter \etal
1999) throughout this paper we adopt a Cold Dark Matter (CDM) cosmological
model with parameters $h=0.7$, $\Omega_0$ = 0.3, $\Omega_\Lambda$ =
0.7, and $\Omega_b = 0.045$, where $h$ is the Hubble constant in units
of 100 km \, s$^{-1}$ Mpc$^{-1}$, and $\Omega_0$,
$\Omega_\Lambda$, and $\Omega_b$ are the total matter, vacuum, and
baryonic densities in units of the critical density, $\rho_{\rm crit} 
= 9.2 \times 10^{-30}$ g/cm$^3$  for our choice of $h$.  Note, however,
that as the outflow-minihalo interactions relevant to GC formation
are all at high redshifts, the value of $\Omega_\Lambda$ has no 
direct impact on our calculations. 

\subsection{The Outflow}

To model the expanding outflow we 
consider a (spherically symmetric)
Sedov-Taylor blast wave with energy input $\epsilon
\E55$ in units of $10^{55}$ ergs, which is expanding into a gas of
$\delta$ times the mean density at a redshift $z$. 
While any realistic starburst-driven outflow will show some
degree of asymmetry, this model gives a resonable approximation
to the structure of high-redshift starbursts seen in 
numerical simluations (\eg Mac Low \& Ferrara 1999;
Mori, Ferrara, \& Madau 2002; Fujita \etal 2004).
Here $\epsilon$ is
the fraction of the total kinetic energy from supernovae ($\E55$)
which is channeled into the galactic outflow. 

In our assumed cosmology, the blast's expansion speed is
\be
v_s = 760
\, \delta_{44}^{-1/2} \, (\epsilon \E55)^{1/2} \,
\left(\frac{1+z_s}{10}\right)^{-3/2} \,  R_s^{-3/2} \qquad \kms,
\label{eq:vs}
\ee
where $\delta_{44} \equiv \delta/44$,
$z_s$ is the redshift at which the shock reaches the halo and
$R_s$ is the physical (ie.\ not comoving) radius of the shock
in units of kpc.
Our choice of the a typical gas overdensity is motivated by the 
model described in \S2.2.
The corresponding postshock temperature is 
$T_s = 1.4 \times 10^5 \, [v_s/100 {\rm km/s})]^2$ K,
assuming an ionized gas with a mean molecular weight of $\mu = 0.6.$
By the time it reaches $R_s$, the shock will have entrained a total mass of
\be
M_{s,{\rm total}}  = 1.3 \times 10^6  
\, \delta_{44} \, \left(\frac{1+z_s}{10}\right)^3 R_s^3 \qquad M_\odot,
\label{eq:Ms}
\ee
of material and have an overall surface density of
\be
\sigma_s = 1.0 \times 10^5 \, \delta_{44} 
\,\left(\frac{1+z_s}{10}\right)^3 \, R_s \qquad \msun \,{\rm kpc}^{-2},
\label{eq:sd}
\ee
and it will reach this radius at a time of
\be
t_s = 0.55 \, \delta_{44}^{1/2} (\epsilon E_{55})^{-1/2} 
\left( \frac{1+z_s}{10} \right)^{3/2} R_s^{5/2} \qquad {\rm Myr}.
\ee

We can estimate the total mass in metals as roughly 2 $\msun$ per
10$^{51}$ ergs, a relation that is true both for type II supernovae,
as well as for pair-production supernovae (SN$_{\gamma\gamma}$) 
from very massive PopIII stars (Woosley \& Weaver 1995; Heger \& Woosley 2002).
Assuming that half of these metals escape from the host galaxy, we
find the mass of ejected metals, in units of $\msun$, to be simply
$M_Z = 10^4 \E55$.   While the Sedov-Tayor solution
assumes that the entrained material represents swept-up gas, in fact
a few times the mass in metals is injected into the blast wave.  However,
the Sedov solution will be accurate as long as
$R_s \gg (\E55/\delta)^{1/3} (1+z_s)^{-1}$ kpc, and cooling within 
the bubble is small.

To estimate the energy in a typical high-redshift outflow, consider
a young galaxy whose total mass is $M_{\rm gal} \ge 10^9 \msun$.  Such
a galaxy will have a virial temperature exceeding 20,000 K, so its
gas readily cools via atomic (HI) line emission, leading to star
formation.  The assumption that 10\% of the gas is converted to stars
gives good agreement with observed high-redshift star-formation rates,
as well as with abundances of metals measured in high-redshift quasar
absorption line systems (Thacker, Scannapieco, \& Davis 2002;
Scannapieco, Ferrara, \& Madau 2002).   For very massive (PopIII)
stars, the models of Heger and Woosley (2002) suggest that supernovae
produce some $10^{51}$ ergs for every 30 $\msun$ of material in new
stars.  For less massive (PopII) stars, a Salpeter IMF yields one
supernova with an explosion energy of $10^{51}$ ergs for every 150
$\msun$ in new stars (eg.\ Tegmark, Silk, \& Evrard 1993).
Thus we obtain 
\be
E_{55} ({\rm PopIII}) \simeq 50 M_9 \qquad {\rm and} \qquad
E_{55} ({\rm PopII}) \simeq 10 M_9 
\ee
where here and below $M_9$  is the galaxy mass in units of $10^9 \msun.$
For the wind efficiency ($\epsilon$)
we rely on the simulations
described in Mori, Ferrara, \& Madau (2002), which indicated $\epsilon
\simeq 0.3$ in the case of a $2 \times 10^8 \msun$ star-bursting galaxy.

Finally, we note that a lower bound, $\epsilon \E55 > 1 $, follows
from the fact that high-redshift outflows dispersed metal efficiently 
enough to preclude stars of primordial composition being formed today
(Scannapieco, Schneider,
\& Ferrara 2003).  This minimum value, plus the numbers cited above,
suggest a fiducial outflow model in which $\epsilon = 0.3$
and $\E55 = 10$.

\subsection{The Protocluster}

Having established basic outflow parameters, we now develop a simple model
for the gas and dark matter in a protocluster whose total mass $M_c =
M_6 \times 10^6 \msun$.  We assume that the gas has a primordial composition
(76\% H and 24 \% He, by mass) and is unionized, giving a mean molecular weight
$\mu = 1.2$.

At a redshift $z_c \sim 10$, corresponding to a cosmic age $\sim 1/2$
Gyr, the gas and dark matter collapse and virialize.  
Initially, the mean density of the protocluster cloud is enhanced by 
a factor $\Delta$ above the background,
\be
\rho_c = \Delta \Omega_0 (1+z_c)^3 \rho_{\rm crit},
\ee
where the enhancement factor for a virialized cloud at high redshift
is well approximated by the value in a critical universe $\Delta = 178$ 
(\eg Eke, Navarro, \& Frenk 1998).
With this choice, the cloud's virial radius is
\be
R_c = 0.3 \, M_6^{1/3} \, \left(\frac{1+z_c}{10}\right)^{-1} \qquad {\rm kpc}.
\ee
and its virial velocity is
\be
v_c = 4.4 \, M_6^{1/3} \, \left(\frac{1+z_c}{10}\right)^{1/2} \qquad \kms.
\ee

As first observed in the numerical simulations of
Navarro, Frenk, and White (1997, hereafter NFW) we assume that the
CDM minihalo develops a radial profile of the form
\be
\rho(R)={\Omega_0 \rho_c \over cx (1+cx)^2} \frac{c^2}{3 F(c)}, \label{eq:NFW}
\ee
where $x\equiv R/R_c$, $c$ is the 
halo concentration parameter, and
\be
F(t)\equiv \ln(1+t)-{t\over 1+t}.
\ee
We then assume that as the procluster gas collapses within the 
dark mater halo, it is shock heated to the virial temperature
\be
T_c = 720 \, M_6^{2/3} \, \frac{1+z_c}{10} \qquad {\rm K},
\ee
and develops the density distribution of isothermal matter in the
CDM potential well:
\be
\rho_{\rm gas}(R)= \rho_0 e^{-\frac{v_{esc}^2(0)-v_{esc}^2(R)}{v_c^2}}.
\label{eq:rhogas}
\ee 
The central gas density ($\rho_0$) is determined by the condition that the 
average baryonic density within the virial radius is equal to 
$(\Omega_b/\Omega_0) \rho_c$.  The escape velocity for an atom at a radius
$R$ within the well is given by
\ba
v_{esc}^2(R = x R_c) & = &2\int_R^\infty {GM_{\rm CDM}(R')\over R'^2}dR' 
\nonumber \\
& = & 2v_c^2\,{F(cx)+{cx\over 1+cx}
\over xF(c)},
\label{eq:ve}
\ea 
so that $v_{esc}^2(0) = 2 v_c^2 c/F(c)$.

Following Madau, Ferrara, 
\& Rees (2001) we assume a typical concentration parameter
of $c=4.8$, although there are some indications that high-redshift
halos may be less concentrated than expected from this estimate
(\eg Bullock \etal 2001).  With this choice $\rho(R_c) = 44 \Omega_b
\rho_{\rm crit}$, motivating our typical value of $\delta_{44}$, and
 $v_{esc}^2(0) = 10 \, v_c^2$, $v_{esc}^2(R_c) = 3.9
\, v_c^2$, and
\be 
\rho_0 ={{178\over 3}c^3\Omega_be^A
\over \int_0^c (1+t)^{A/t}\,t^2 dt} \rho_{\rm crit} = 1970 \rho_{\rm crit},\,
\ee
where $t \equiv cx$ and $A \equiv 2c/F(c) = 10.3$. 

Note that a gas-to-star conversion efficiency of $10 \%$ followed by a
loss of $50 \%$ of a young globular cluster's original stellar mass
(through winds, tidal forces, etc.), when combined with the
aforementioned {\em present} GC mass limit, $M_\star < 10^6 \msun$,
implies that the likely limiting mass of proto-globular clouds
(gas+CDM) is
\be
M_6({\rm protocluster}) < 10^2.
\label{eq:jon11}
\ee
For a collapse redshift $z_c \sim 10,$ the corresponding limit on the
virial temperature is $T_c < 15,000$ K. Choosing a  $20 \%$ 
conversion efficiency gives a corresponding limit on the virial 
temperature of $T_c < 9,000$ K, which is low-enough to preclude 
substantial ionization, hydrogen line cooling, and ``unassisted'' 
star formation.  Thus, given the uncertainties involved, the maximum 
size of minihalos corresponds to the maximum size of halo globular 
clusters.

Lately, a great deal of attention has been given to the inner profile
of dark-matter halos, as measured from the properties of low surface
brightness (LSB) galaxies, whose gravitational potentials  are thought
to be dark-matter dominated at all radii.  In particular, studies of
these objects have found that  their inner rotation  curves are likely
to be well-described by constant density cores, in contrast to the
$1/R$ profile assumed here (\eg Carignan \& Beaulieu 1989; Carignan,
\& Sancisi 1991;  de Blok \& McGaugh 1997).  Recent studies have
complicated the issue, however, suggesting that to some degree, these
results may have been affected by  the poor angular resolution of the
H I observations.  By accounting for beam smearing, several groups
have shown that the  H I rotation curves of LSB galaxies are
consistent with a wide variety of dark matter potentials, ranging from
constant density cores to profiles as steep as  $R^{-1}$ (\eg
Blais-Ouellette \etal 1999; Swaters, Madore, \& Trewhella 2000; van
den Bosch \& Swaters 2001).  Further high-resolution $H\alpha$
measurements have reached similar conclusions (Swaters \etal 2003).
While still an unsettled issue, the presence of such a core will only
reduce the gravitational potential in the center of the cloud,
enhancing the impact of a cloud-outflow interaction.
This effect is only important in the interior $R < R_c/c$ profile of
the cloud,  however, which contains $\sim 15 \%$ of the gas mass.
Thus we expect our NFW model to provide a reasonable description of
the the protocluster; at worst it provides a slight underestimate of
the impact of shocking by galaxy outflows.

Finally we note that, given our cosmological model (with a primordial
power spectrum as given by Eisenstein \& Hu 1999, $\sigma_8 = 0.87$)
and the efficiencies assumed above,  by $z_c = 15$ the region that
evolved into the Milky Way was filled with $\sim 1000$ minihalos  that
could form into GCs with stellar masses $\geq 10^5 \msun$; by $z_c =
10$, there were $\sim 3000$ such minihalos.  These estimates are based
on the Lacey \& Cole (1993)  progenitor model and an assumed total
Galactic mass and formation redshift of $2 \times 10^{12} \msun$ and
$z_f = 2$.  Even assuming just the mean cosmological density results
in $\sim 500$ and $\sim 2000$ such minihalos at $z_c = 15$ and $z_c =
10$ respectively (Press \& Schechter 1974).  Thus, although we show
below that only a fraction  of minihalos will be located the correct
distance from a starburst to form a GC, our mechanism need not be
particularly efficient to account for the Galaxy's $\sim 100$ halo
globular clusters.

\section{Fate of the Protocluster}

\subsection{Outflow-Protocluster Interactions}

Using the simple outflow and protocluster models described above, 
we now outline the general features to be expected when the shock 
interacts with the gas in a CDM minihalo.  The mass of the shell that
impacts such a protocluster is 
\be
M_s = \pi R_c^2 \sigma_s = 2.9 \times 10^4 \, \delta_{44} \, M_6^{2/3} 
  \frac{(1+z_s)^3}{10 (1+z_c)^2} R_s \,\,\, \msun,
\ee
and at the time of impact its momentum is
\ba
P_s =  \pi R_c^2 p_s  = & 2.2 \times 10^7 \, 
M_6^{2/3} \, (\delta_{44} \,\epsilon \, \E55)^{1/2} \nonumber \\
&  \times \frac{10^{1/2}(1+z_s)^{3/2}}{(1+z_c)^{2}}  \, R_s^{-1/2} \,\,
M_\odot \, \kms,
\label{eq:ps}
\ea
where $p_s$ is a momentum surface density.
Recalling that $M_Z = 10^4 \E55 \msun$ is the total mass in metals 
within the blast-wave material, it is straightforward to calculate
that the metallicity to which the protocluster gas is enriched is 
\be
Z_c =  0.076 \, \xi \, M_6^{-1/3} \, E_{\rm 55} \, 
\left(\frac{1+z_c}{10} \right)^{-2} 
\, R_s^{-2}  \qquad Z_\odot,
\label{eq:Z}
\ee
where we have taken solar metallicity to be $1/50$ by mass, and
have assumed that a fraction $\xi$ of the metals is thoroughly mixed 
into protocluster gas and is contained in the GC stars observed today.
This mixing efficiency is studied in detail in \S 4.

We are most interested in clouds whose stars are
enriched to approximately 1/30 the solar value (Ashman \& Zepf 1998), 
which occurs at a typical distance of
\be
R_s = 1.5 \, M_6^{-1/6} \, (\xi \E55)^{1/2} \, \left(\frac{1+z_c}{10} \right)^{-1} 
\qquad {\rm kpc}.
\label{eq:rmetal}
\ee
This is sufficiently large that the original mass input is negligible
compared to the entrained mass (see eq.\ \ref{eq:Ms}).
The time it take for the bubble to reach this radius is
\ba
t_s = & 1.5 \, M^{-5/12} \, \delta_{44}^{1/2} \,\epsilon^{-1/2} 
\, \xi^{5/4} \, E_{55}^{3/4} \,  \nonumber \\  &  \times 
\left(\frac{1+z_s}{10} \right)^{3/2} 
\left(\frac{1+z_c}{10} \right)^{-5/2} \, Z_{1/30}^{-5/4} 
\,\,\, {\rm Myr}.
\ea
Typically, this is shorter the cooling time within the bubble
\be
t_{\rm cool} = \frac{\frac{3}{2} n \, k T_s}{n^2 _e\Lambda} 
= 18 \, T_6 \,  \delta_{\rm 44}^{-1} \, \left(\frac{1+z}{10} 
\right)^{-3} \Lambda_{-23}^{-1}  
\,\,\, {\rm Myr},
\label{eq:tcool}
\ee
where $\Lambda_{-23}$ is the radiative cooling rate of the gas in
units of $10^{-23}$ ergs cm$^3$ s$^{-1}$ and $n_e$ is electron number
density of the shocked gas in units of cm$^{-3}$,
$T_6$ is the gas temperature in units of $10^6$ K,
and we have asssumed that the density of the post-shock gas
is enhanced by a factor of 4.
Thus our assumed Sedov solution should be reasonable for the
range of values considered in this study.

Futhermore, for likely values of $\E55$, $R_s$
is comparable to the virial radius of the outflowing galaxy,
which is $30 M_9^{1/3} (1+z_s)^{-1}$ kpc. Hence, the density of the
medium through which the shock is passing is substantially higher than
the mean IGM density, and more like our fiducial value,  $\delta = 44$,
taken at the virial radius. 

From Eqs.\ (\ref{eq:vs}), (\ref{eq:Ms}), (\ref{eq:ps}), and (\ref{eq:rmetal}) 
we determine the shock velocity at the time of impact to be
\be
v_s = 420 \, 
(\delta_{44}^{-2} \,M_6 \, \epsilon^2 \, \xi^{-3} \, \E55^{-1})^{1/4} 
\left( \frac{1+z_c}{1+z_s} \right)^{3/2} \, \kms,
\label{eq:vs2}
\ee
and the mass and momentum impinging on the cloud to be
\be
M_s = 4.3 \times 10^4  \, (\delta_{44}^2 \, M_6 \, \xi \, \E55)^{1/2}
\left( \frac{1+z_s}{1+z_c} \right)^{3} \,\,\, M_\odot, 
\label{eq:Ms2}
\ee
and
\ba
P_s &=& 1.8 \times 10^7 \, 
(\delta_{44}^2 M_6 \, \epsilon^2 \, \xi^{-1} \, \E55)^{1/4} \nonumber \\
& & \times \left( \frac{1+z_s}{1+z_c} \right)^{3/2} \qquad M_\odot \, \kms.
\label{eq:ps2}
\ea
If we compare $P_s$ with the virial velocity of the cloud times 
its baryonic mass,
\be
P_c = 6.3\times 10^5  M_6^{4/3} \left(\frac{1+z_c}{10}\right)^{1/2} \qquad
M_\odot \, \kms,
\label{eq:muc}
\ee
we see that for redshifts $z_c \approx z_s \approx 10$ the blast wave's 
momentum is sufficient  to move the gas
out of the dark matter potential well whenever $M_6 \leq 10^2
\left( \frac{\epsilon^2 E_{55}}{\xi} \right)^{3/7}$. But, how will such
a dark cloud be impacted as a function of radius, and what is the
ultimate fate of its swept-up gas?

\subsection{Three Stages of Evolution}

In order to study this interaction in more detail, we adopt a specific set of
parameters for \S3.2, and in \S3.3 we discuss the effect of varying these
values.  Here we put $E_{55}$, $\epsilon = 0.3$, $z_c = 10$, $z_s = 8$,
and $Z_c = 1/30 Z_\odot$.  
We consider a protocluster with 
$M = 3.2 \times 10^6 \msun$ whose total gas mass is
approximately $5 \times 10^{5} \msun$.
Assuming an overall star formation efficiency $\sim 10\%,$ this provides
a good match to the observed peak in the globular cluster mass function at
$10^5 M_\odot$ (in stars).  In this section, 
we assume  that all metals eventually 
mix into the protocluster gas, so $\xi = 1$.  Altogether, these choices
represent our fiducial model.

\begin{figure}
\centerline{\psfig{figure=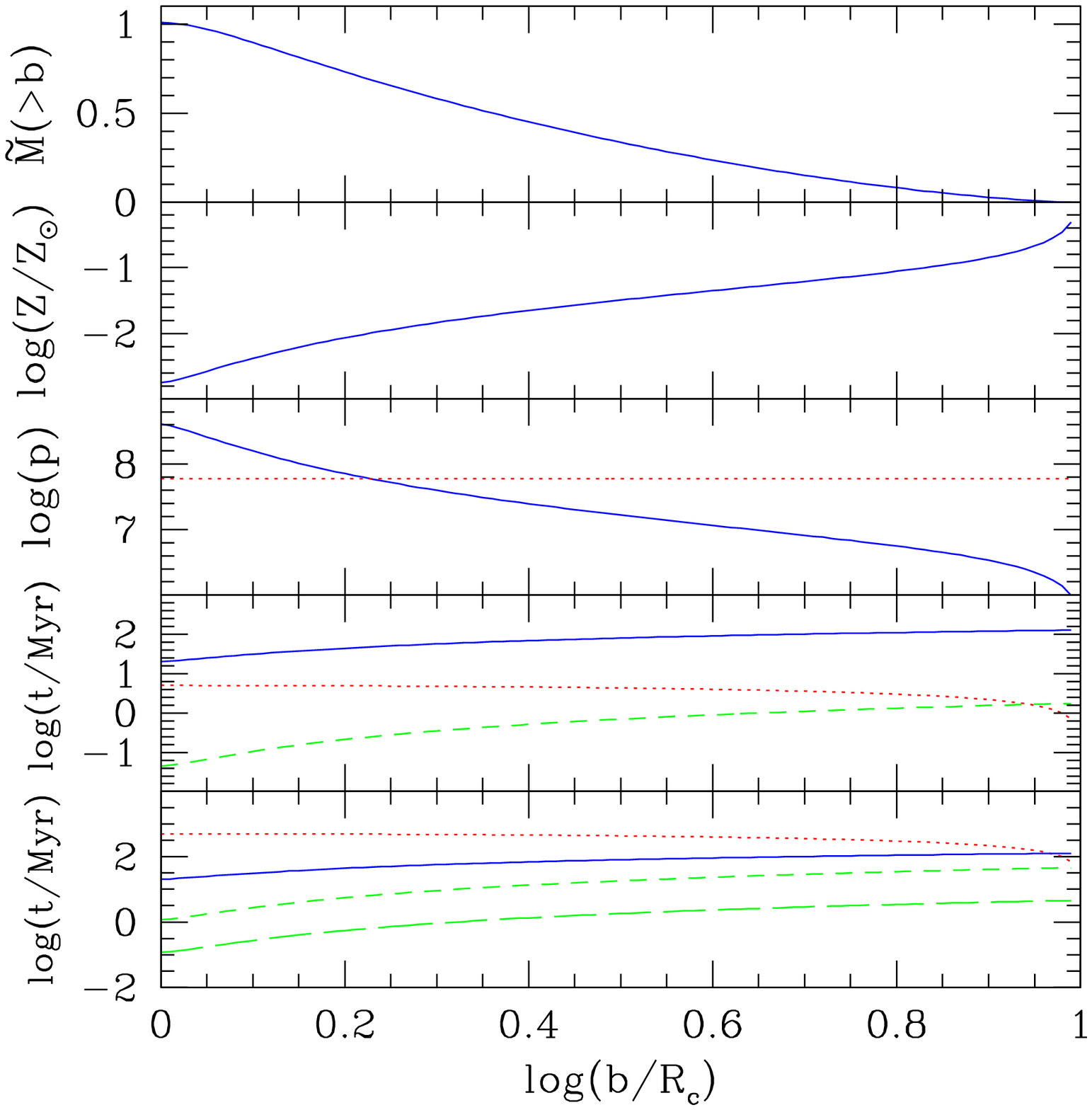,height=17cm}}
\caption{Fiducial model of globular cluster formation ($\E55 = 10$,
$\delta = 44$, $\epsilon = 0.3$, $M = 10^{6.5} M_\odot$
$\log(Z_c/Z_\odot) = -1.5$).  {\em Top panel:} fractional gas mass
($\tilde{M}(>b)\equiv M(>b)/M_{\rm gas}$) outside a given impact
parameter.  {\em Second panel:} Local metallicity as function of $b$.
{\em Third panel:} Cloud momentum surface density, $p_c(b)$, (solid
line) versus impinging momentum per unit area (dotted line), both
plotted in units of $\msun \, \kms \, {\rm kpc}^{-2}.$ {\em Fourth
panel:} Comparison of the sound-crossing time (dotted), free-fall time
(solid), and cooling time (dashed) just after the shock has passed
over the minihalo.  {\em Fifth panel:} Comparison of the sound-crossing
time, free-fall time, and cooling time after the cloud
has cooled to $100$K.  Lines are as in the fourth panel, while the
long-dashed line is the cooling time at 1000K.}
\label{fig:j0}
\end{figure}

We identify three
important stages in the evolution of shocked cloud.  The first of
these occurs as the outflows moves across the minihalo.  At this
point, the key question is whether the impinging momentum is
sufficient to accelerate the gas to its escape velocity, stripping it
from the dark matter potential.  To estimate when this occurs, we
compute the average momentum surface density as
\be
p_c(b) = 2 \, \int^{\sqrt{R_c^2-b^2}}_0 \, d\ell \,
\rho_{\rm gas}(\sqrt{b^2+\ell^2})\,
v_{esc}(\sqrt{b^2+\ell^2}),
\ee
$b$ is the impact parameter from the central axis of the cloud,
and $\ell$ is the distance along a line parallel to this axis.  
This is plotted in third row of Fig.\ \ref{fig:j0},
for our fiducial model.

Comparing the momentum surface density  with the momentum per unit
area in the outflow, we find that $p_s > p_c(b)$ for all impact
parameters  $b \gsim \bstrip = 0.2 R_c.$  In our simple picture, we
assume that all the gas outside of $\bstrip$ will be stripped from the
potential, while the denser central regions will be left behind,
resulting in an elongated bell-shaped distribution, not unlike the
coma of a comet.

In order to better relate our radial profiles
to the properties of the cloud, in the
upper two panels of Fig.\ \ref{fig:j0} we plot 
\be
M(>b) =  2 \pi \, \int^{\sqrt{R_c^2-b^2}}_0 \, db' \, \sigma_c(b'),
\ee
the mass outside a given impact parameter, and
\be
Z_c(b) = \frac{Z_c M}{\pi R_c^2 \sigma_c(b)},
\ee
the local metallicity at that distance,
where $\sigma_c(b) \equiv 2 \int^{\sqrt{R_c^2-b^2}}_0 dz 
\rho_{\rm gas}(\sqrt{b^2+z^2})$.  From these estimates, we find
that  approximately $80\%$ of the gas mass is able to be efficiently
stripped from the minihalo and that, once homogenized,
 the metallicity of this gas exceeeds $10^{-2} Z_\odot.$

The second important stage occurs just after the blast front 
passes across the protocluster.  At this time, the gas has been heated to a
postshock temperature of several million degrees and 
its density has been enhanced by a factor of four.  There are
three time scales that then enter into the problem.  The
first of these is the sound crossing time, which we estimate 
as 
\ba
t_{\rm sc} &\equiv& \frac{(R_c^2-b^2)^{1/2}}{4 c_s} \\ &=& 
6.8 \, T_6^{-1/2} M_6^{1/3} \, (1+z_c)^{-1} \left[1-(b/R_c)^2 \right]^{1/2}
\, {\rm Myr}, \nonumber
\ea
where $c_s = 0.43 v_s = 115 T_6^{1/2}$ km/s is the postshock sound speed.

The first key issue is then whether 
the self-gravity of the {\em gas alone} can resist the pressure
associated with such an enormous temperature increase. This is
determined by comparing the sound crossing time with the free-fall 
time, which can be written in convenient units
as
\be
t_{\rm ff} = \sqrt{\frac{3 \pi}{32 G \rho_{\rm gas}}} = 
67 \, n^{-1/2} \qquad {\rm Myr},
\ee
where $G$ is the gravitational constant, 
and $n$ is the density of the gas in atoms cm$^{-3}$.
In the fourth panel of Fig.\ \ref{fig:j0} we see that for our fiducial
set of parameters $t_{\rm ff} \gg t_{\rm sc}$ at all impact parameters.  
Thus it seems the swept away gas will be evaporated into the IGM within 
a sound crossing time.
In order for this to occur, however, the gas must expand before cooling
process are able to dissipate its thermal energy.
This is determined by the cooling time as given by eq.\ (\ref{eq:tcool}).
Note that in this equation
$\Lambda_{-23}$ is a function not only of the temperature of the
gas, but also of its elemental composition, as the presence of heavy
elements greatly increases the number of transitions that
can radiate efficiently.  For simplicity we estimate this
radiation by assuming that all the gas is at the mean metallicity given
by eq.\ (\ref{eq:Z}) and by taking solar abundance ratios,
which allows us to make use of the tabulated models of Sutherland \&
Dopita (1993). This approximation is dependent on the prompt 
mixing of the minihalo gas with all the impinging material (as discussed
in \S4), but in fact 
a range of metallicities may be found in the gas at a given
impact parameter.  For temperatures within  $10^7 {\rm K} \gtrsim T 
\gtrsim 10^4,$ this leads to $\Lambda_{-23}(Z = 1/30 Z_\odot)$ 
values ranging from 0.6 to 10. 
At higher temperatures, cooling is dominated by bremsstrahlung
and is largely metallicity independent, while molecular cooling
becomes important at lower temperatures, as discussed below.
Finally we estimate the post-shock density at an impact
parameter $b$ as $\sigma_c(b)\left[(R_c^2 - b^2)^{1/2}/4 \right]^{-1}.$

With these simplifying assumptions, we obtain 
the cooling times that are plotted as the dashed lines
in the fourth panel of Fig.\ \ref{fig:j0}.  Note that
$t_{\rm cool}$ is several orders of
magnitude smaller than the free-fall time,
as can be inferred directly from the properties
of GCs observed today (Murray \& Lin 1992).
In fact, for a large range of impact parameters, $\tcool$ is much
smaller than the sound crossing time. Thus, despite enormous
postshock temperatures, efficient radiation by the dense and
metal-enriched halo gas is able prevent evaporation of the gas within
$b \leq 0.7 R_c,$ which we label as $\bcool$.  This means that roughly
half of the gas ($0.2 R_c \leq b \leq 0.7 R_c$) is expelled from the
dark matter potential, yet cools sufficiently quickly to remain
gravitationally bound.  Modulo our assumption of 
efficient mixing ($\xi \sim 1$),  this gas has a relatively small
range of metallicities, from about $10^{-2}$ to $10^{-1.5}$ $Z_\odot$.

At high temperatures, the cooling time is a monotonic function that
strongly decreases with decreasing temperature.  The cloud becomes
ever more efficient at radiating its energy until the gas begins to
cool below $10^4$K, when it reaches the third and final important
stage.   At this point, the gas is largely neutral, and the cooing
rate decreases precipitously.  Two new processes then become
important, the first of these is the production of molecular hydrogen
by nonequilibrium reactions (\eg Ferrara 1998) and the second is
infrared line emission by \CII, \FeII, and \SiII, whose ionization
potentials are less than  13.6 eV (Dalgarno \& McCray 1972).   In the
case of molecular hydrogen,  runaway cooling from $\sim 10^6$K results
in appreciable levels of H$^-$ and H$^+_2$, which act as
intermediaries in H$_2$ formation.  While the exact numbers are
uncertain to within a factor $\sim 3$ this can result in levels of
H$_2$ $\sim 1\%$  (see \eg Uehara \& Inutsuka 2000, Figure 1), which,
in turn yield cooling  rates of $\sim 10^{-26} T_3^{2.5}$ ergs cm$^3$
s$^{-1}$ (Galli \& Palla 1998).   The atomic infrared  cooling is
somewhat less efficient but also a weaker function of temperature,
ranging from about $10^{-27}$ ergs cm$^3$ s$^{-1}$ at 1000 K to
$10^{-28}$ ergs cm$^3$ s$^{-1}$ at 100 K for $1/30 Z_\odot$ gas. Thus
cooling through $H_2$ is expected to be slightly more important than
metal line cooling at 1000 K and slightly less important at 100 K.

The relevant times scales at late times are plotted in the bottom
panel of Fig.\ \ref{fig:j0}, assuming that the gas cools at a fixed
density.  Here we see that at both 1000 and 100K, the cooling time is
substantially shorter than the sound crossing time at all radii,  
and the gas is likely to cool to very low temperatures even if
photo-dissociating radiation (which we do not attept to model)
were strong enough to quickly
destroy the formed $H_2$, meaning that all cooling was through 
\CII, \FeII, and \SiII, 
or if mixing is inefficent, meaning that all cooling was through
$H_2$. In fact, at 100 K the sound crossing times even lie above the free-fall 
times.  Thus, although the cooling rate is orders of
magnitude smaller at these low temperatures, weak radiation from $H_2$
or metals is still sufficient to cool the cloud to the point at which
its thermal pressure cannot counteract self-gravity.   Note that this
runaway collapse takes place even if we assume no additional density
enhancement during cooling, and no external pressure from the hot
medium that is likely  to be found behind the shell [as seen for
example in NGC 3077 (Ott, Martin, \& Walter  2003)]. As $t_{\rm ff}
\propto n^{-1/2}$ and $\tcool \propto n^{-1}$, while $t_{\rm sc}
\propto R \propto n^{-1/3}$, such density changes during cooling from
$\sim 10^6$K will only enhance collapse at low  temperatures.
Furthermore external pressure, while negligible for the hot cloud, may
equal or even exceed the thermal pressure of $\sim 1000$K gas.  Thus
we believe low-temperature cooling does not represent an important 
barrier for star formation in shocked minihalos.

To summarize, the two key issues that determine the fate of the gas are
momentum transfer to the cloud during outflow shocking (stage 1)
and the ratio of the cooling time to the sound crossing time  {\em
just after} the shock moves across the cloud (stage 2).  It is  these
issues that we now address in some detail, as we study the impact of
varying several model parameters about their fiducial values.

\subsection{Dependence on Input Conditions}

\begin{figure}
\centerline{\psfig{figure=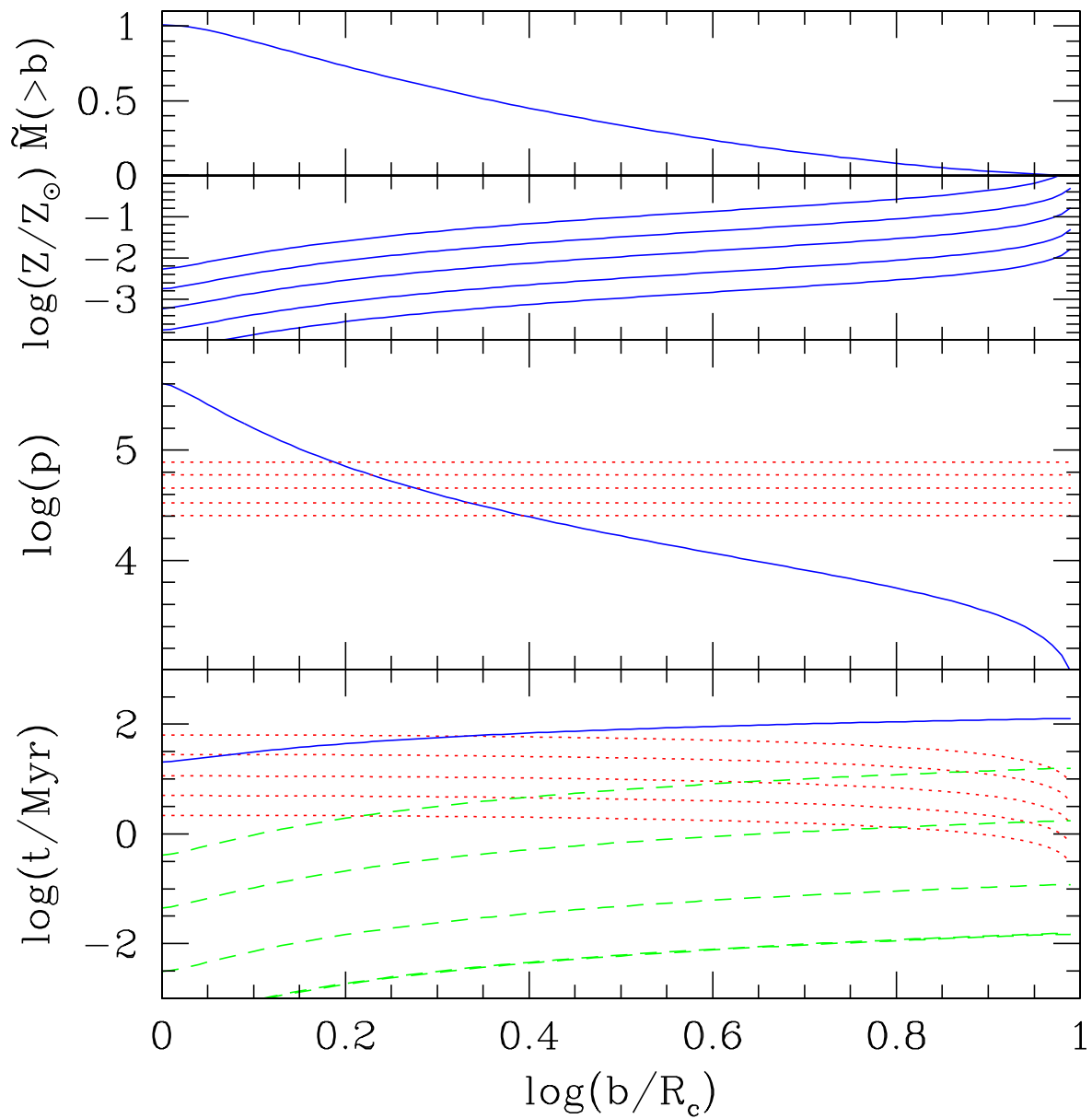,height=17cm}}
\caption{Effect of varying the distance from the outflowing
galaxy. Distances are chosen such that the overall mass average
metallicities are  $\log(Z_c/Z_\odot)$ = -3.0, -2.5, -2, -1.5, and
-1. Curves are as in the top four panels of Fig.\
\protect\ref{fig:j0}.  As metallicity increases, the shock reaches the
minihalos with more momentum and energy, increasing both $p_s$ (dotted
lines, center panel) and $\tcool$ (dashed lines, bottom panel).  As
the cloud is shocked to higher temnperatures, increasing $Z_c$
decreases the sound  crossing time.  Minihalo clouds are disrupted in
the highest metallicity case, while shocking is too weak to trigger
star formation in the lowest-metallicity case.}
\label{fig:jZ}
\end{figure}

Having outlined the general stages we expect shocked minihalos to
undergo, we now consider a range of values for the input parameters of
our model, and relate these to the observed range of properties of halo
globular clusters.   The first issue we explore is the impact of
varying the distance between the minihalo and the outflowing galaxy.
As this is directly related to the mass in metals reaching the cloud,
according to eq.\ (\ref{eq:Z}), this constrains the range of globular
cluster metallicities that can be generated in our model.  
Holding all other fiducial paramters fixed,
we select mean $Z_c$ values of
$10^{-3.0} $, $10^{-2.5} $, $10^{-2} $, $10^{-1.5} $, and $10^{-1}
Z_\odot ,$ which correspond to distances $R_s$ of 21, 12, 6.6,  3.7, and
2.1 kpc, respectively, and to the results that  are plotted in
Fig.\ \ref{fig:jZ}.

At the smallest distance, more momentum is imparted to the halo,
causing stripping to be slightly more efficient.  As the momentum
surface density is proportional $R_s^{-1/2}$ however, this effect is
minor in comparison to the change in the postshock temperature $T_s
\propto v_s^2 \propto R_s^{-3}.$ Thus, in this case, $\tcool$  is
vastly increased while $t_{\rm sc}$ is decreased, and the stripped gas
is evaporated without forming stars.  Similarly, increasing the
distance has a much stronger impact on $t_{\rm sc}$ and  $\tcool$ than
on $p_s$.  In the models in which $Z =10^{-2}$,  $10^{-2.5}$, and
$10^{-3} Z_\odot$, almost all of the gas outside of $b \approx 0.3$,
$0.35$ and $0.4 R_c$ respectively is stripped from the halo, yet able
to efficiently cool, which again corresponds to roughly $50 \%$ of the
gas mass.  In fact the maximum distance at which ejection and
shock-induced star formation are effective is not determined by the
stripping criteria, but rather by the postshock temperature being too
low to ionize the minihalo gas and allow it to radiate efficiently.
The post-shock temperature for $Z_c = 10^{-3} Z_\odot$, for example,
is only $\sim 5000$K, well below the $\sim 10^4$K  needed for
effective collisional excitation of hydrogen.

These results compare well with the observed metallicity distribution
of halo globular clusters, which peaks slightly below
[Fe/H$] = -1.5$ and falls off gradually with very few objects having iron
abundances below [Fe/H$] = -2.5$ or above [Fe/H$] =  -1$ (Ashman
\& Zepf 1998).  Furthermore the
relatively constant metallicity between $0.2 R_c \lesssim b \lesssim 0.7 R_c$
compares well with the observed $\Delta[{\rm Fe/H}] \leq 0.1$ dex
spread observed within individual objects (\eg Suntzeff 1993).

\begin{figure}
\centerline{\psfig{figure=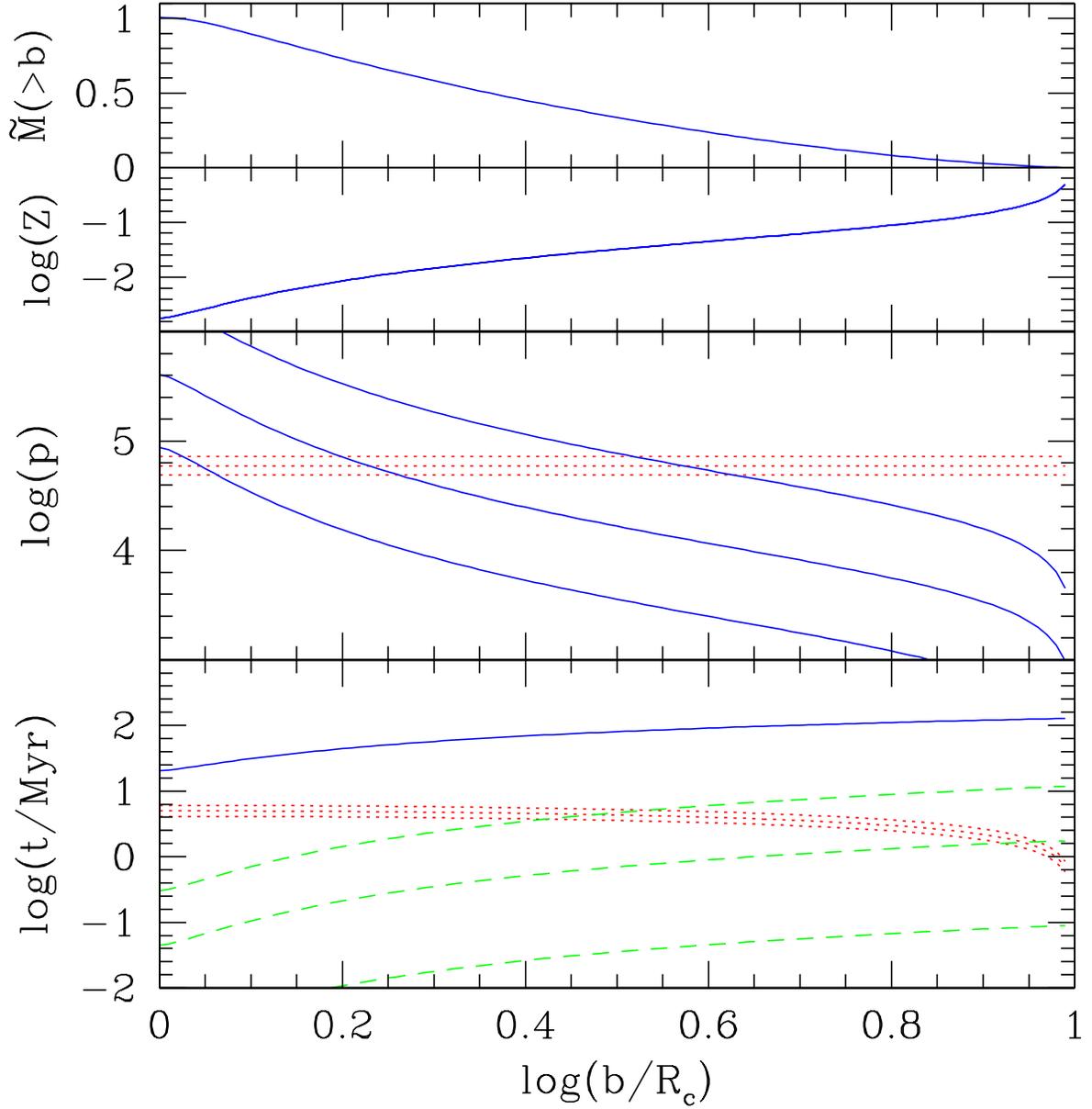,height=17cm}}
\caption{Effect of varying the minihalo mass.  Panels are as in 
Fig.\ \protect\ref{fig:jZ}, in all cases curves
are for masses of $10^{5.5} \msun$, $10^{6.5} \msun$,
and $10^{7.5} \msun$.  
Increasing the mass while fixing $Z_c$ raises $p_c$ (solid lines,
center panel) dramatically 
and moves the minihalo closer, increasing the impinging
momentum surface density to a lesser extent (dotted lines, center panel).  
Similarly  increasing the 
minihalo mass increases the cooling time dramatically
(dashed lines, lower panel)
and although the cloud is hotter, the increase in mass causes the sound
crossing time to increase slightly.  
Thus shock induced star formation is most efficient
in lower-mass clouds, while long cooling times suppress star formation
in the largest minihalos.}
\label{fig:jM}
\end{figure}

The second important issue is the dependence of our model on the mass
of the shocked cloud.  In Fig.\ \ref{fig:jM} we vary this quantity over
two orders of magnitude, while keeping other fiducial parameters fixed.
By eq.\ (\ref{eq:Z}),
this means that the distance between the source galaxy and the minihalo
is assumed to increase with decreasing mass.  Thus, although stripping is
seen to be more efficient in the lower mass case, the post-shock temperature is
reduced too, as $T_s \propto v_s^2 \propto M_c^{1/2}$ at a fixed
$Z_c$.
This greatly reduces the cooling time, allowing the gas as a whole
to be stripped and to cool quickly enough to remain bound.

From this comparison it is clear that our assumed mechanism  becomes
more effective in the case of smaller minihalos, and that the minimum
scale of globular clusters probably depends instead on processes that
occur after star formation.  Fortunately, as discussed in \S 1, this
limit is easily understood in terms of the dominant destruction
processes that act on GCs in the Milky Way, in particular
disk-shocking and mechanical evaporation.  On the other hand, the
maximum mass of GCs is likely to be a feature of the proto-cluster
cloud itself.   Turning our attention to the high mass case, we
find that if $M_c = 10^{7.5} M_\odot$ then $\bstrip = 0.5 R_c$ and
$\bcool = 0.25 R_c.$ Thus we expect our scenario to be relatively
inefficient in forming stars in objects more than a few times larger
than our fiducial mass, $M_c = 10^{6.5} \msun$,
which rougly corresponds to the largest minihalo as well
as the maximum observed GC mass, as discussed in \S2.2.

\begin{figure}
\centerline{\psfig{figure=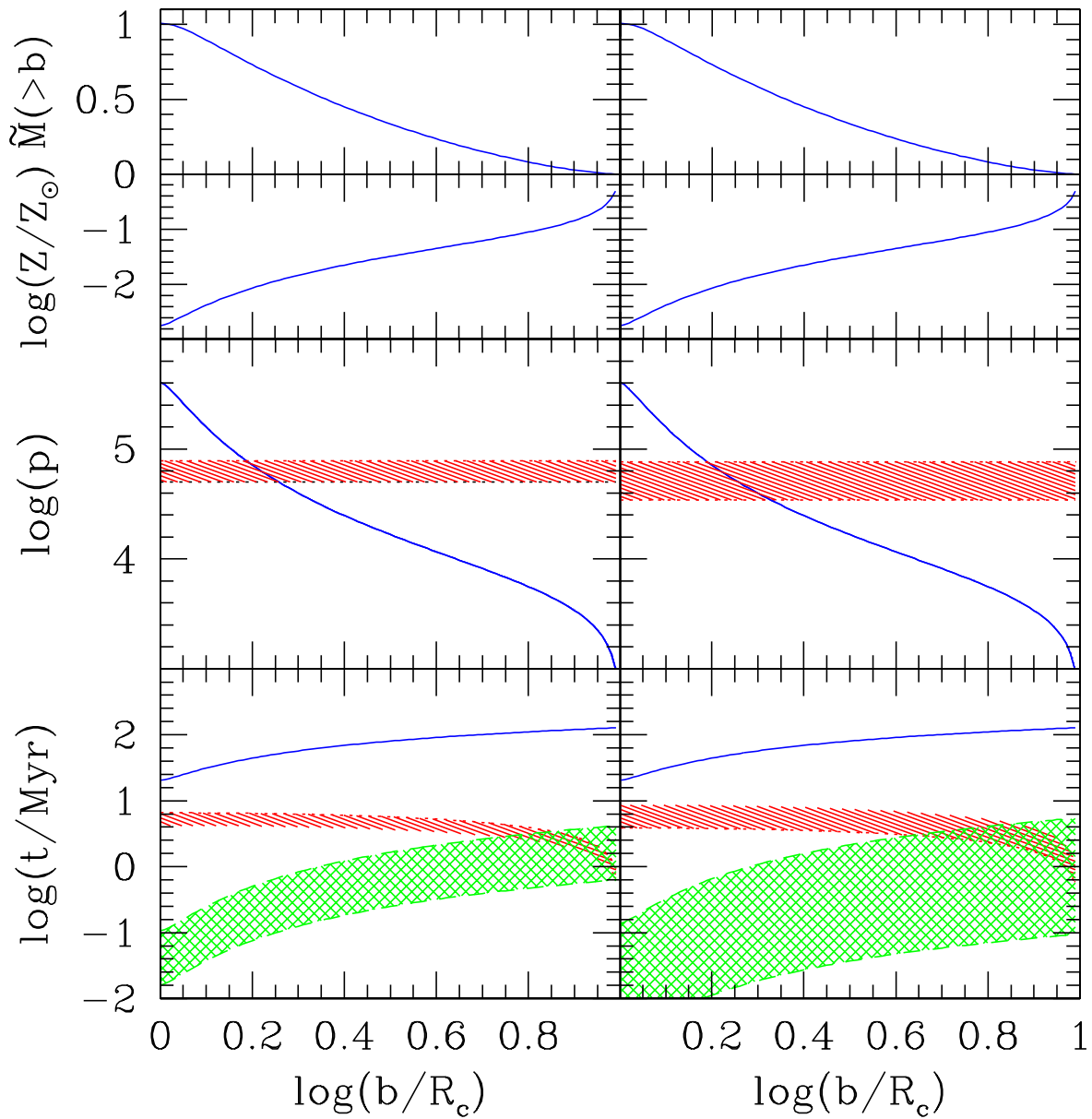,height=17cm}}
\caption{Effect of varying the energy input ($\E55$) and the wind
efficiency ($\epsilon$).  Panels as in Fig.\ \protect\ref{fig:jM}.
{\em Left Column:} $\E55$ is varied between 5 and 30, keeping $Z_c$
fixed.  The largest value of $\E55$ corresponds to the largest
impinging momentum surface density 
in the center panel (shaded region), but to the
{\em lowest} post-shock temperature.  Thus $\E55 = 30$  corresponds to
the largest value of $t_{\rm sc}$  (shaded region bottom panel)
and the smallest value of $t_{\rm cool}$ (cross-hatched region bottom
panel).  This is because objects of a fixed metallicity  lie at
larger distances as $\E55$ is increased.  {\em Right Column:}
$\epsilon$ is varied between .1 and .5.  As with $\E55$, an increase
in the efficiency parameter increases $p_s$ (shaded region, center
panel), but in this case $t_{\rm sc}$ decreases (shaded region bottom
panel), and $t_{\rm cool}$ goes up (cross-hatched region bottom panel).}
\label{fig:jE}
\end{figure}

Next we evaluate the effect of varying the energy input parameters
$\E55$ and $\epsilon$.  Increasing $\E55$ increases the mass of
ejected metals, so raising this value also raises the distance
corresponding to $10^{-1.5} Z_\odot$ enrichment.  Thus, changing the
number of supernovae in the outflowing galaxy has a relatively minor
impact on the momentum imparted to the cloud, changing $\bstrip$
by only a few percent, as is shown in the left
panels of Fig.\ \ref{fig:jE}.  Similarly, simultaneously increasing
the energy input and distance serves to damp the effect $\E55$
has on the sound crossing time and and cooling time of the postshock gas.  Note
that as $R_s \propto E^{1/2}$ the postshock temperature is proportional
to $R_s^{-3} E \propto E^{-1/2}$, {\em decreasing} slightly with
increasing $\E55$, and shorting $t_{\rm cool}$ while raising $t_{\rm
sc}$ somewhat.  Thus the primary impact of $\E55$ on our globular
cluster formation model is simply to shift the relevant distance
further away from the outflowing galaxy, while keeping the
metallicity fixed,  and slightly improving the
fraction of minihalo gas available for star formation.

In the right columns of Fig.\ \ref{fig:jE} we study the impact of 
varying $\epsilon$, the ejection efficiency.  As this parameter affects only
the shock velocity, it has no impact or our choice of fiducial
distance and is not subject to the damping observed in the left
panels.  In this case $\bstrip$ ranges from 0.2 $R_c$ to 0.35 $R_c$ 
while $\bcool$
ranges from $0.4 R_c$ in the $\epsilon = 0.5$ case to $R_c$ if
$\epsilon = .1$.  Thus the wind efficiency represents a larger model
uncertainty than $\E55$, although at least 40\% of the gas mass lies
between $\bstrip$ and $\bcool$ in all cases.

\begin{figure}
\centerline{\psfig{figure=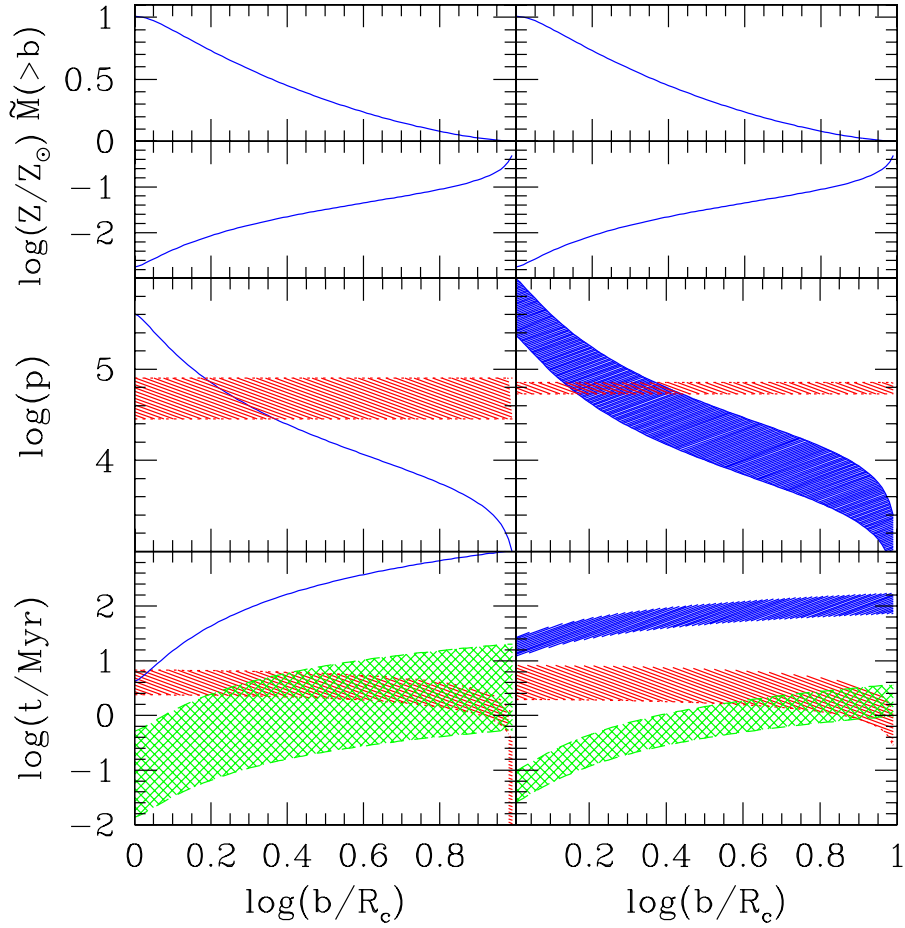,height=13cm}}
\caption{Effect of varying the density surrounding the outflowing
galaxy ($\delta$) and the cloud virialization redshift ($z_c$),
panel and curves as in Fig. \protect\ref{fig:jE}.
{\em Left Column:} $\delta$ is varied between 10 and 80. Increasing
 $\delta$ increases $p_s$  and reduces 
the post-shock temperature.  
Thus as $\delta$ goes up,
$\tcool$ decreases (cross hatched region, lower panel), 
and $t_{\rm sc}$ goes up (shaded region, lower panel), improving the
formation efficiency.
{\em Right Column:} $z_c$ is varied between 8 and 15. This has essentially
the opposite effect from $\delta$.  Thus increasing $z_c$ decreases $p_s$,
increases $\tcool$, and reduces $t_{\rm sc}$, making our mechanism less
efficient.   Finally, increaing $z_c$ also results in a more compact
cloud, with a higher escape velocity (solid region, center panel)
and a shorter free-fall time (solid region, bottom panel).}  
\label{fig:jd}
\end{figure}

Finally, in Figure \ref{fig:jd} we study the effect of varying the
density of the medium through which the shock is moving, as determined 
by $\delta$, and the density of the minihalo, as determined by $z_c$.
From eqs.\ (\ref{eq:vs2}) and (\ref{eq:ps2}) $p_s \propto \delta^{1/2}$
while $T_s \propto \delta^{-1}.$   Thus increasing $\delta$ improves
the efficiency of our formation mechanism, pushing $b_{\rm strip}$ 
towards the center, while shifting $b_{\rm cool}$ outwards.  Similarly,
decreasing $\delta$ reduces both the efficiency of stripping and cooling.
At $\delta = 10$, $b_{\rm cool} = 0.15 R_c < b_{\rm strip} = 0.35 R_c$
and our GC formation mechanism fails completely. Thus,
our scenario is most likely to take place at
separations that are comparable to the virial radius of the
starbursting galaxy.  
This point is particularly noteworthy because, while protocluster-scale
minihalos are ubiquitous in CDM cosmological simulations, no patently 
intergalactic globular clusters have been found to date (McLaughlin 1999;
Mar\' in-Franch \& Aparicio 2003).

Changes in the density of the cloud, as determined by its formation
redshift, have the opposite effect on the shock properties.  In this
case $p_s \propto (1+z_c)^{-3/2} \propto \delta_c^{-1/2}$ and $T_s
\propto (1+z_c)^3 \propto \delta_c.$   Variations in $z_c$ also effect
the properties of the cloud, however, in particular shortening the
free-fall time and increasing the imparted momentum $p_c$ as the
formation redshift is increased.  Thus later-forming clouds are
heavily favored by our  mechanism, and is unlikely that efficient
star-formation was achieved in objects in which $z_c \gg z_s.$  Note
that it is the ratio of $1+z_s$  and $1+z_c$ that is important for
this comparison, and thus we expect little difference in our results
if both the collapse and outflow redshifts were shifted to higher
values, as might be necessary if  cosmic reionization took place
early, as  suggested by polarization measurements from the {\em
Wilkinson Microwave Anisotropy Probe} (Kogut \etal 2003).

To summarize these sensitivity studies in the context of our scenario, 
cloud-outflow interactions are efficient at
forming stars in clusters with mean metallicities ranging from about
$10^{-2.5} Z_\odot$ to $10^{-1} Z_\odot$.  Below $10^{-2.5} Z_\odot$ 
shocks are too weak to induce star formation, while above 
$10^{-1} Z_\odot$, shocks are too powerful and disrupt the
gas completely.  Induced star formation is largely independent of the 
total energy input from a given starburst, although some uncertainty
is introduced by the unknown fraction of this energy that goes into
powering a galaxy outflow.  While outflowing shells are efficient at
forming stars in small minihalos, this become more difficult in the
largest of such objects, whose gas masses compare well with the maximum
stellar masses of globular clusters.  And finally, our mechanism
is most efficient in cases in which the density contrast between the
cloud and the surrounding medium is not too high: in minihalos
close to starbursting galaxies that formed at a similar redshift.

\section{Turbulence Models of Metal Mixing}

\subsection{A 1D Code for Turbulence Transport and Mixing}

In \S3  we saw that the chemical homogeneity observed within globular
clusters can be associated with the range of metallicities 
generated in outflow-minihalo interactions.  Yet these estimates
were based only on the mass of heavy elements impinging on the
cloud as a function of impact parameter.  For true homogeneity within
the resulting stellar population to be achieved, not only must
sufficient masses of metal be present, but this material must be well 
mixed into the star-forming gas.  Note that this chemical structure is 
of particular importance as it plays a key role in gas cooling as
described in \S3.2.

As a first step towards examining this ``fine grained'' mixing  (Rees
2003) in detail, we have developed a one-dimensional numerical code,
which make use of a well-tested generalization (Besnard \etal 1992)
of the the widely used ``$K-\epsilon$'' model of turbulence transport
(Harlow \& Nakayama 1967).  In this case the equations of
conservation of mass, momentum, and  energy are coupled to the
Reynolds stress tensor, $R_{i,j} \equiv \overline{\rho u''_i u''_j}$
which represents the ensemble average of the  product of the density
and the velocity departures from the mass-weighted mean. Thus we have
\ba
\frac{\partial \rho}{\partial t} +
\frac{\partial \rho u_1}{\partial x_1} &=& 0, \\
\frac{\partial \rho u_i}{\partial t} +
\frac{\partial \rho u_i u_1}{\partial x_1} &=&
-\frac{\partial P}{\partial x_1}
-\frac{\partial R_{i,1}}{\partial x_1},\\
\frac{\partial \rho E}{\partial t} +
\frac{\partial \rho E u_1}{\partial x_1} &=& 
-\frac{\partial P \bar u_1}{\partial x_1}
-\frac{\partial R_{i,1}}{\partial x_1} \nonumber \\
& & +\frac{\partial}{\partial x_1}
\left(  \frac{\rho \nu_t C_p}{{\rm Pr}_t} 
\frac{\partial T}{\partial x_i}\right),
\ea
where $t$ and $x_1$ are time and position variables,
$\rho(x_1,t)$ is the mass density field, $u_i(x_1,t)$ 
is the mass-averaged mean-flow velocity field in the $i$ direction,
$P(x_1,t)$ is the mean pressure, $E(x_1,t)$ is the total mean energy,
$T(x_1,t)$ is the mean temperature, $\bar u_i(x_1,t)$ is the volume
averaged mean-flow velocity field in the $i$ direction, and we take 
${\rm Pr}_t = 1.0$ and $C_{\rm p} = 3/2$, as appropriate for
a polytropic gas.
The Reynolds stress tensor is then modeled 
using a Boussinesq-type approximation as
\be
R_{i,j} = -\rho \nu_t 
\left( \frac{\partial u_i}{\partial x_j} +
       \frac{\partial u_j}{\partial x_i} \right)
+ \frac{2 \rho}{3} \delta_{i,j} 
\left( \nu_t \frac{\partial u_k}{\partial x_k} + K \right),
\ee
where $\nu_t$ is the turbulence ``eddy'' viscosity (which is
calculated as $\nu_t = 0.09 k^2 \epsilon^{-1}$), $K(x_1,t)$ is the turbulence
energy density, and $\epsilon(x_1,t)$ is the turbulence energy density
decay rate. The transport equation for each of these quantities is 
constructed 
as a compressible generalization of the standard $K-\epsilon$ model, namely
\be
\frac{\partial \rho K}{\partial t} +
\frac{\partial \rho K u_1}{\partial x_1} =
a_1 \frac{\partial P}{\partial x_1}
- R_{i,1}\frac{\partial u_i}{\partial x_1}
- \rho \epsilon
+\frac{\partial}{\partial x_1}
\left(  \rho \nu_t \frac{\partial K}{\partial x_i}\right),
\label{eq:k}
\ee
where $a_i \equiv \overline{\rho' u_i'}/\rho$
(with $\rho'$ and $u'$ the 
departures from the volume-weighted mean density and velocity),
and
the terms on the right hand side of the equation represent
buoyancy creation, shear creation, decay, and self-diffusion
respectively.
Similarly 
\ba
\frac{\partial \rho \epsilon}{\partial t} +
\frac{\partial \rho \epsilon u_1}{\partial x_1}&=& 
\frac{\epsilon}{K} \left[
C_{\epsilon 4} a_1 \frac{\partial P}{\partial x_1}
- C_{\epsilon_1} R_{i,1}\frac{\partial u_i}{\partial x_1}
- C_{\epsilon_2} \rho \epsilon \right] \nonumber \\ 
& & +\frac{\partial}{\partial x_1}
\left(  \frac{\rho \nu_t}{\sigma_\epsilon} 
\frac{\partial \epsilon}{\partial x_i}\right)
-C_{\rm \epsilon 3} \rho \epsilon \frac{\partial u_1}{\partial x_1},
\label{eq:epsilon}
\ea
where the first four terms on the right hand side parallel those
in eq.\ (\ref{eq:k}), and the final term represents changes in
the turbulent scale due to expansion or contraction of the
fluid element.  This scale can be defined as
\be
S \equiv \frac{K^{3/2}}{\epsilon},
\ee
such that an expanding motion results in an increase in $S$,
which translates to a decrease in $\epsilon$,
likewise a contraction decreases $S$, driving $\epsilon$ up.
In eq.\ (\ref{eq:epsilon})
$\rho a_i$ is the turbulence mass flux, while the empirically
fit constants are    
$\sigma_\epsilon = 1.3$,
$C_{\epsilon_1} = 1.55$,
$C_{\epsilon_2} = 2.0$,
$C_{\epsilon_3} =  \frac{1}{d} + 1 -\frac{2}{3} C_{\epsilon_1}$,
and $C_{\epsilon_4} = 1.25$,
(\eg Launder, Reece, \& Rodi 1975)
where $d$ is the dimensionality of the compression modeled in
the last term of eq.\ (\ref{eq:epsilon}), which in our case is
1.  Closure of our system
of equations is achieved by the transport equation for $a_i$:
\be
\frac{\partial \rho a_i}{\partial t} +
\frac{\partial \rho a_i u_1}{\partial x_1} =
- \frac{R_{i,1}}{\rho} \frac{\partial \rho} {\partial x_1},
- C_{a1} \frac{\rho \epsilon}{K} a_i + \frac{b_{\rm trb}}{\rho} 
\frac{\partial P}{\partial x_1},
\ee
where the first two right-hand terms represent turbulence distributive 
creation and ``drag'' decay, $C_{a1} = 2.0$, and the last term
with $b_{\rm trb} 
\equiv \overline{\rho' \rho'}/\rho$, describes buoyancy effects,
which are here neglected.
Mixing between multiple materials is computed by carrying additional
transport equations for their concentrations, $c_\alpha$.  For this 
purpose we use a diffusional approximation for the species flux:
\be
\frac{\partial \rho c_\alpha}{\partial t} +
\frac{\partial \rho c_\alpha u_1}{\partial x_1} =
\frac{\partial}{\partial x_1}
\left(  \frac{\rho \nu_t}{\sigma_c} 
\frac{\partial c_\alpha}{\partial x_i}\right)
\label{eq:calpha}
\ee
where $\alpha$ is an index over materials and  $\sigma_c = 1.3$.
We compute $\bar u_i$ as $u_i - a_i$ and  
the temperature as 
$T = \frac{2\mu m_p}{3 k_B}\left(E - K - \frac{u^2}{2} \right),$
with $k_B$ the Boltzmann constant.
Our code does not attempt to account for self-gravity or
the impact of the dark matter gravitational potential, as these 
only become of primary importance when the full two-dimensional
geometry of the distortion is calculated, as discussed in \S4.2.

This formulation of turbulence equations is closely related to that
used in the CAVEAT numerical code (Adessio \etal 1990),  where it has
been compared with a wide range of experimental results.  Our
implementation is an Eulerian finite-difference one, with donor-cell
advection being used throughout.  Test calculations show overall
qualitative agreement  with experimental results for the shear
problems of interest us.   While our approach is sufficiently accurate
to demonstrate our ideas for this paper, more precise analyses may
require the use of a full spectral representation, like that described
by Steinkamp, Clark, \& Harlow (1999a; 1999b).

\subsection{Application to Shocked Minihalos}

\begin{figure}
\centerline{\psfig{figure=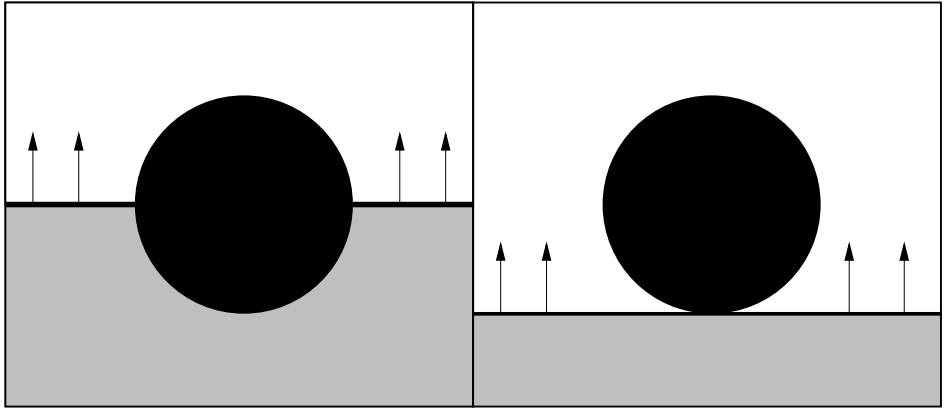,height=6cm}}
\caption{Illustration of the edge-on (left panel) and face-on (right 
panel) configurations.  In both panels the cloud of primordial composition
is indicated in black, while the impinging enriched material is
indicated in grey.}
\label{fig:simsketch}
\end{figure}

Having developed and tested a simple one-dimensional code to track
turbulent mixing in compressible fluids, we then applied this tool  to
the problem of cloud-outflow interactions.  Here we focused on two key
geometries: an edge-on configuration, in which mixing occurs due to
Kelvin-Helmholtz  instabilities as the outflow shears along the side
of the cloud, and a face-on configuration, in which mixing due to
Rayleigh-Taylor  instabilities occurs as the outflow material moves
into the denser cloud.  Each of these configurations is illustrated in
Figure \ref{fig:simsketch}.

Turning our attention first to the shear instability case, we chose our
initial conditions to approximate the fiducial problem described in
\S3.  Our simulation domain was divided into 800 evenly-spaced zones,
spanning a distance of 840 physical parsecs, twice the radius
of our fiducial $M =  10^{6.5} M_\odot, z_c = 10$ cloud. 
The first
400 of our simulations 
zones were filled with material 1, which was taken to
represent the primordial gas making up the
cloud.  Here the density  was computed as per eq.\
(\ref{eq:rhogas}), the fluid was initially at  rest, and the
temperature was taken to be the virial value of 1650 K.  The second
400 zones were filled with material 2, which was taken to
represent the enriched outflow.  In these cells the density was taken
to be 4 times the density at the radius of the cloud, $u_y$ 
was 200 km/s ($\sim 200$ parsecs/Myr), and the temperature 
was taken to be the postshock value of
$560,000$ K.  On both sides of the shock, material is assumed to be
largely quiescent, with an initial turbulent kinetic energy equal to
only $2 \%$ of the total energy.  An initial turbulent scale $S$ of
10 parsecs was chosen in both materials.

\begin{figure}
\centerline{\psfig{figure=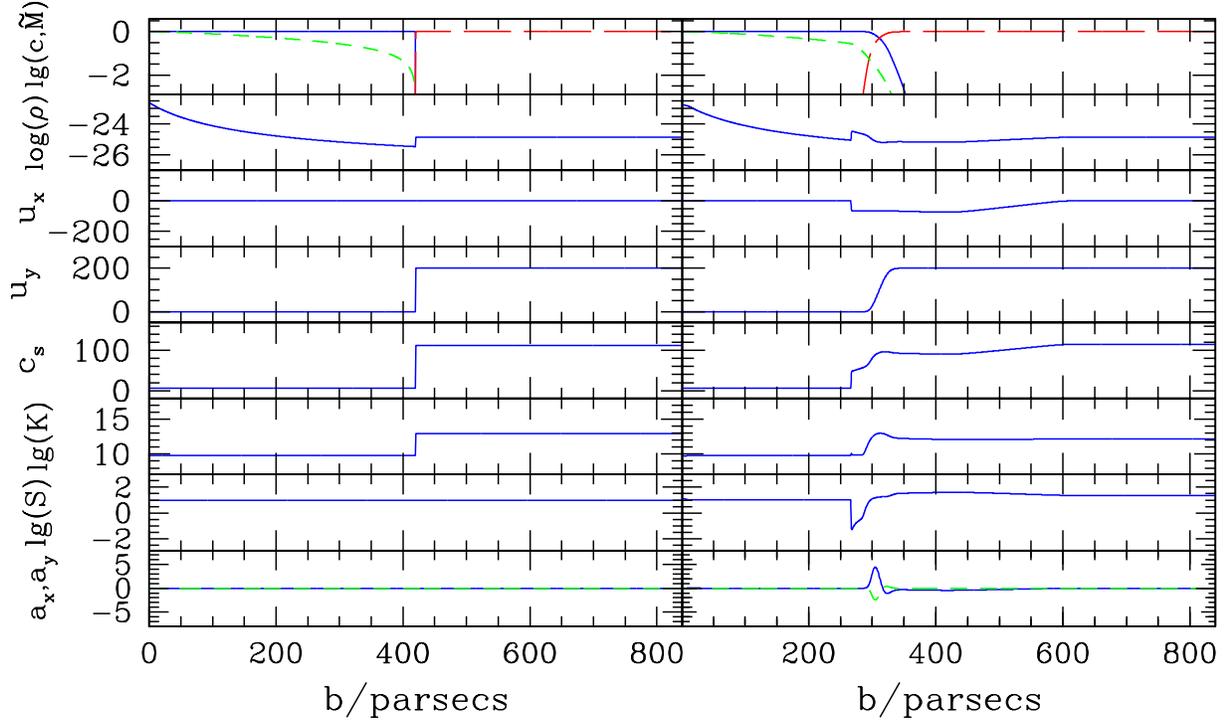,height=9cm}}
\caption{Quantities in the Kelvin-Helmholtz shear problem at times
$t=0$ (left column) and $t=1$ ($1.5 \times 10^6$ years, right column).  In the
top row the solid line is $c_1$, the long-dashed line is $c_2$, and
the short-dashed line  is a measure of the normalized cloud mass $\tilde M(b)$
exterior to an impact parameter, $b$.
In the second row, the units of $\rho$ are  grams per
cm$^3$.   In the third through fifth rows $u_x$, $u_y$, and $c_s$
are  in km/s.   The last three rows represent turbulent quantities.
Among these $K$ (sixth row), is in ergs/gram, $S$ (seventh  row)
is in units of parsecs, and $a_x$ and $a_y$ (represented by the solid
and dashed lines in the final row) are in km/s.   The initial
turbulence kinetic energy is assumed to be 2\% of initial total
energy.}
\label{fig:shearturb0}
\end{figure}

These conditions are shown in the various panels in the left 
column of Figure 
\ref{fig:shearturb0}.  In order to better quantify mixing
in the simulation, we also computed 
\be
\tilde M(b) \equiv 1 - \frac{\int_0^b dx c_1(x) \rho(x) x^2}
                      {\int_0^{R_c} dx'' c_1(x') \rho(x') x'^2},
\label{eq:Mb}
\ee
a quantity that estimates the fraction of the total cloud gas mass 
contained outside of  a given  impact parameter $b$, 
which is plotted as the dashed  lines in the upper panel.

\begin{figure}
\centerline{\psfig{figure=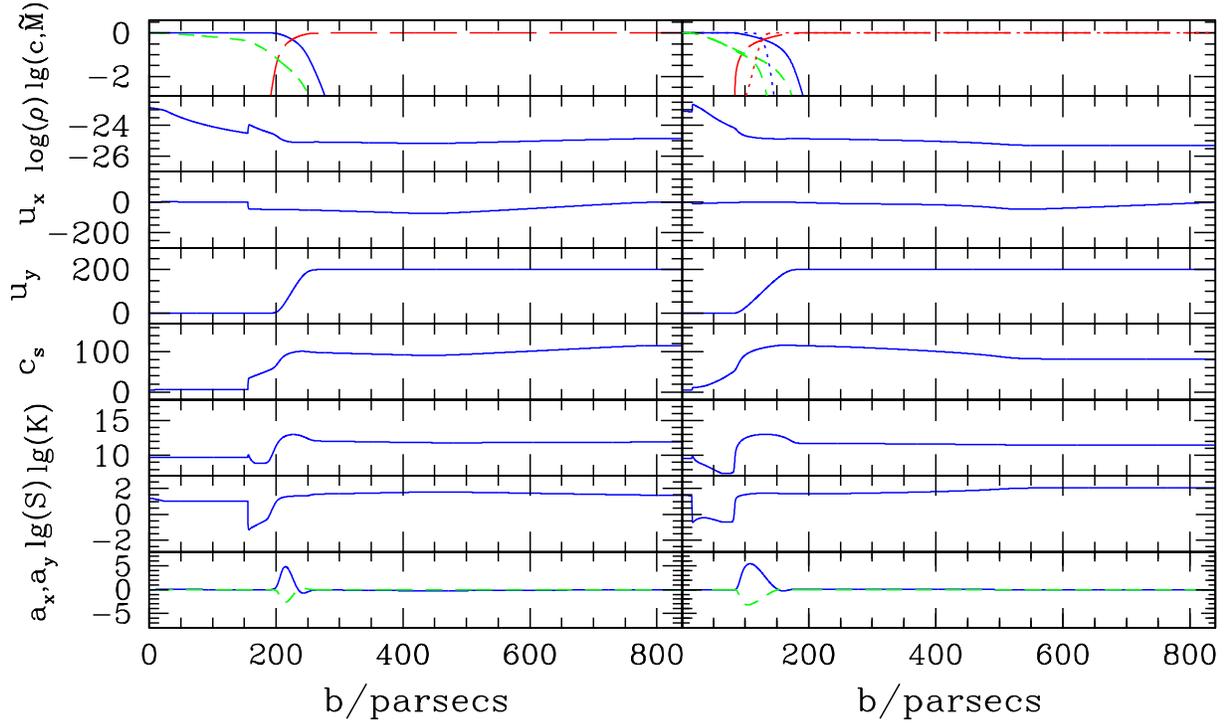,height=9cm}}
\caption{Kelvin-Helmholtz shear problem at times $t=2$
($3.0\times 10^6$ yr, left column) and $t=5$ 
($7.5 \times 10^6$ years, right column).  
Rows are as in Figure  \protect\ref{fig:shearturb0},
and the additional dotted lines 
in the uppermost $t=5$ panel show
$c_1$ and $c_2$ in a case in which $\nu_t$ has been set to zero
in the transport equations for the concentrations, and the lower
dashed curve shows $M(\ell)$ for this case.}
\label{fig:shearturb2}
\end{figure}

Adopting a time step of $600$ years and a time unit  such
that $t = 1$ is defined as $R_c/v_s = $ 1.5 million years, we
ran our simulation to a  final time of $t=5$.  From the
estimates in $\S3.1$, one can expect a $\sim 20$ Myr delay between
shocking and star formation, which is primarily governed by the free-fall
time; however by these late times the distortion of the cloud is likely
to be so severe that our one-dimensional calculations will
look nothing like the true configuration.  Thus we choose to stop
our simulation much sooner, quoting our simulated mixing as a lower
limit.  These results are depicted
in Figures \ref{fig:shearturb0} and \ref{fig:shearturb2}.   

Three major features are visible in the mean-flow quantities
represented in the upper panels in these plots:  a strong shock moving
into the cloud, a material interface moving inwards, 
and a rarefaction wave moving outward into the enriched
gas.   The motion of each of these features is well approximated by
the analytical solution for the propagation of a strong pressure
discontinuity in the absence of turbulence transport or shear (eg.\
Harlow \& Amsden 1971):
\ba
u_{x,int} &\approx&
- 3 (c_{s,{\rm outflow}} - c_{s,r}),\\
u_{x,s} &\approx& \frac{4}{3} u_{x,int}, 
\qquad{\rm and} \label{eq:usuint}\\
u_{x,r} &\approx&  c_{s,{\rm outflow}},
\ea
where  $u_{x,{\rm int}},$ $u_{x,{\rm s}},$ and $u_{x,r}$ are  the
velocities of the interface, shock, and rarefaction wave respectively,
$c_{s,{\rm outflow}}$ is the sound  speed in the outflow region, and
$c_{s,r}$ is the sound  speed to the left of the rarefaction wave.
Initially, the inward shock and fluid interface move at speeds of
$\sim 90$ and $\sim 65$ km/s respectively, but slow somewhat as they
reach the dense regions near the center of the cloud.  On the other 
hand, the sound speed in the unshocked regions is quite small ($\sim 5$
km/s).   As the cloud is assumed to be initially in pressure
equilibrium, any motions due to gravitational effects should also be
$\sim 5$ km/s, justifying our neglecting self-gravity and the 
dark-matter gravitational potential in this one-dimensional calculation.

Turning our attention to the turbulence energy profile, we find
a notable increase along the shearing interface
between the two materials.  This is initially
confined to a relatively narrow region at early times, but 
expands to cover $\sim 100$ parsecs by $t=5$, 
and the corresponding turbulent diffusion
is sufficient to mix metals into $\xi \sim 20\%$ of the cloud by
this time.  In order to distinguish this mixing from the numerical 
diffusion intrinsic to our Eulerian approach, we also include $c_\alpha$
values in the right column of Figure \ref{fig:shearturb2} for an identical run
in which $\nu_t$ has been set to zero in eq.\ (\ref{eq:calpha}).

As a test of convergence, both runs were repeated doubling 
the resolution and halving the  overall time step.  This
had the effect of reducing the mixing in the $\nu_t = 0$ estimates 
of the concentrations, while leaving other quantities unchanged in 
both runs.

\begin{figure}
\centerline{\psfig{figure=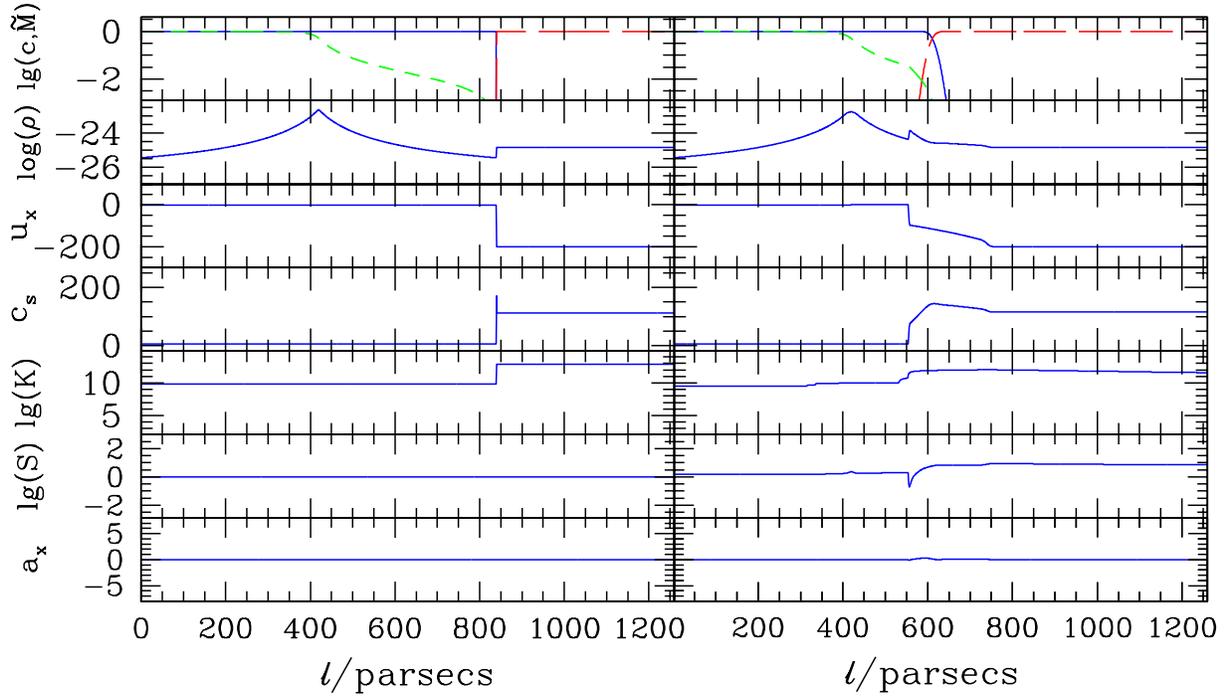,height=9cm}}
\caption{Face-on Raleigh-Taylor problem at 
times
$t=0$ (left column) and $t=1$ ($1.5 \times 10^6$ years, right column).  
In the upper row the solid line is $c_1$, the long-dashed line is
$c_2$, and the short-dashed line  is
the normalized cloud mass $\tilde M(\ell)$
contained in front of a given distance from the back
of the cloud, $\ell$.
Rows 2-4 represent mean flow quantities, and rows 5-7 represent
turbulent quantities, with units as is \protect\ref{fig:shearturb0}.}
\label{fig:faceonturb0}
\end{figure}

To explore our model in the face-on case, we conducted a
second simulation with initial conditions as shown in the left column
of Figure
\ref{fig:faceonturb0}.  Here the simulation domain was divided into three
regions, the left two representing the full density profile through
the cloud, and the right representing the impinging outflow.   The
simulation domain was divided into 1200 cells, each with  the same
width that was used in the edge-on calculation ($1.05$  parsec).
Again the dashed lines in the upper panels represent the total
mass of primordial material contained in front of a given $\ell$ value, 
$\tilde M(\ell)$, 
where $\ell$ is defined as the distance from the back of the cloud.
This mass 
is calculated using an equation analogous to eq.\
(\ref{eq:Mb}), but now adding over the full cloud such that $\tilde M =
0.5$ at the center of the density profile.

\begin{figure}
\centerline{\psfig{figure=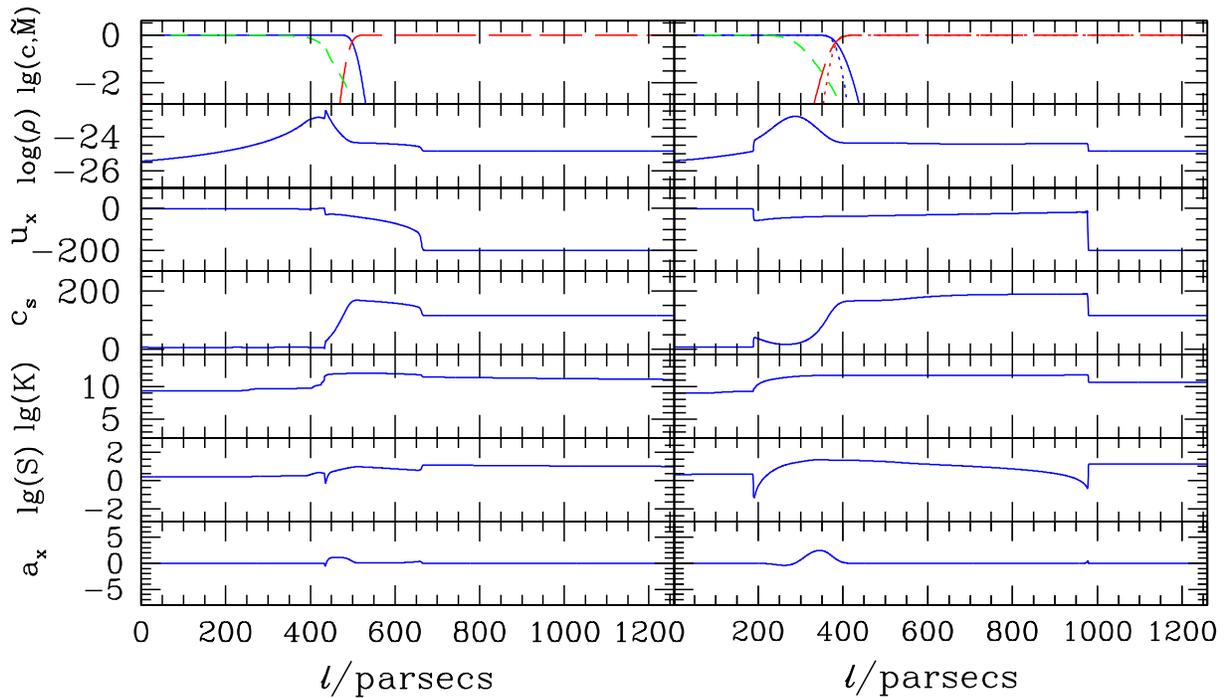,height=9cm}}
\caption{Face-on Raleigh-Taylor problem at times $t=2$
($3.0\times 10^6$ yr, left column) and $t=5$ ($7.5\times 10^6$ yr, 
right column).  
Rows are as in Figure 
\protect\ref{fig:faceonturb0}, and, as in Figure \protect\ref{fig:shearturb2},
additional dotted lines have been added in the upper right panel which show
$C_1$ and $c_2$ in a case in which $\nu_t$ has been set to zero in the
transport equations for the concentrations.}
\label{fig:faceonturb2}
\end{figure}

Choosing a time step and time unit as in the edge-on case, we ran our
simulation to a final time of $t=5$,  with results as depicted in
Figures \ref{fig:faceonturb0} and \ref{fig:faceonturb2}.  Again, three
major features are visible in the mean-flow quantities.  As in the
edge-on case, a strong shock and material interface quickly propagate
inwards.   Initially, the fluid interface moves at the incoming
speed of $200$ km/s, and $u_{x,s}$ and $u_{x,int}$ share roughly the
relation expected from eq.\ (\ref{eq:usuint}).  But as in the
Kelvin-Helmholtz problem, both of these features are  slowed
considerably as they move into the central regions of the cloud,
where the gas density becomes large.

Unlike the edge-on case, the third major feature is not a
rarefaction wave, but rather a reflected shock, which gains
in strength as the inward propagating shock moves into the 
increasingly dense material.  Thus this feature
initially moves leftward, at a physical velocity of $\sim - 100$ km/s,
but a positive velocity $\sim 100$ km/s in the frame of the
incoming shock.  By $t = 5$ however, this shock attains
a positive physical velocity $\sim 80$ km/s, and exhibits
a density ratio $\sim 4.$ 

In the lower three panels of these figures, we again find
an increase in the turbulent kinetic energy along the interface,
but with a lower overall magnitude than in the shear layer case.
In fact, the most efficient source of $K$ in this simulation
is the backwards moving shock, which is made up completely
of enriched gas.    This means that mixing in this geometry
is less efficient than in the edge-on case, and 
our results indicate that only
$\xi \sim 3\%$ of the total cloud mass is able to be enriched before
our final time step.
Comparing this mixing to an identical run in which 
$\nu_t$ was set to zero in eq.\ (\ref{eq:calpha}), we find
that much of this mixing is due to numerical diffusion.
At least in the one-dimensional case, mixing is dominated by
shear, and occurs at a level $\xi \gtrsim 20 \%$ before the 
onset of star formation.

From these calculations we expect a reasonable value for the
fraction of impinging metals mixed throughly into the cloud to be
$\xi \sim 50\%$.   Furthermore, even higher $\xi$ values may be
appropriate when the motion of the flow is not constricted to such
limited geometries, as hinted at by a feature in our face-on calculations.
In the regions between the material interface and the reverse shock, 
a large turbulence scale length is generated by $t \sim 1$ and grows 
with time, suggesting that a global instability may be developing within
the cloud. In fact, while not strictly turbulence,
a global mean-flow distortion is known to arise
in an analogous configuration in the laboratory, where  it
serves as a prime example of the problem of transition to turbulence.
In the experimental studies described in Vorobieff \etal (2003), a
column of heavy gas (SF$_6$) initially  surrounded by air was
accelerated by a planar shock.  In this case the cloud is observed
to deform as an arc, which buckles
along the edges.  Eventually these edges spiral over themselves,
leading to two large counter-rotating vortices as shown in Figure 
\ref{fig:experiment}.  These vortices quickly 
transition to an overall turbulent flow, unlike the one
seen in our one-dimensional studies.

It is this global distortion that is likely to be key in
increasing mixing in the post-shock cloud.   Indeed, a similar
geometry was hinted at by the arcs observed in the 
jet-cloud interactions studied by
Fragile \etal (2004), as well as by the
the  large-scale vortices observed in the shock-cloud simulations of
Klein, McKee, \& Colella (1994).  However the properties of this overall
buckling will undoubtedly be modified by the presence of a radial
density gradient, forces due to the dark-matter potential, and self
gravity, none of which were included in these studies.
Clearly then, for the moment, we can only conclude that the simple
Kelvin-Helmholtz and Rayleigh-Taylor instabilities simulated in our
one-dimensional calculations are insufficient to
produce the observed homogeneity within globular clusters.  The
question of global mixing requires more extensive modeling, an issue
we are actively pursuing and intend to describe in a future
publication.

\section{Discussion}

The first cosmic structures to form are likely not to have been
galaxies, but rather lower mass clouds of gas and dark matter  too
small to cool by atomic processes.  In fact, in the Cold Dark Matter
cosmological model assumed here, as much as $15 \%$ of the total
mass is contained in such minihalos by a redshift of 10.
Thus these clouds represent a vast reservoir of material that can
quickly be converted into stars though shocking and enrichment by
neighboring objects.

A second generic feature of  any hierarchical  model is a
high-redshift epoch of galaxy outflows.  Although the sizes of
galaxies increase strongly with time,  internal properties such as the
scale of OB associations  (McKee \& Williams 1997) and the efficiency
of energy deposition from supernovae are largely independent of redshift.
Thus while the ejecta of conglomerations of Type II SNe are primarily
confined to the interstellar medium of large galaxies like the Milky Way 
forming at $z \sim 2$, they are easily able to escape from the dwarf
galaxies forming at $z \gtrsim 3$, generating  large outflows of
material such as are now well-observed in the Lyman break population.

It is clear that  these outflows will interact with minihalos, 
the uncertainty lies only in the nature of that
interaction.  Through our estimates in \S3, we have shown that for a
large range of outflow energies, minihalos sizes, and separations, the
most likely outcome is a dense cluster of stars that is bounded by
self-gravity, but stripped from its associated dark matter.
Furthermore the present-day universe has natural candidates for 
these collections of stars, which formed on the periphery of galaxies.

Although globular cluster formation may well be an ongoing  process,
the majority of {\em halo}  globular star clusters are old, and exhibit
several  features of induced minihalo star-formation.
Observations of GC tidal streams show no indication of
associated dark mater halos, unlike the accreted dwarf spheroidal galaxies
(\eg Ibata \& Lewis 1998; Mayer \etal 2002).  GCs also exhibit a
maximum mass $\sim 10^6 M_\odot$ that can not be understood by
dynamical friction or any other known destruction mechanism that
operates in the Galaxy.  Instead, this value closely corresponds 
to the minimum gas mass of high-redshift  minihalos for which
atomic cooling at the virial temperature is effective.
Finally, the span of time $\Delta t$ over which halo GCs 
form in our model is fairly short, on the order the parent galaxy's 
virial radius divided by its outflow velocity.  
Even for a large galaxy like the Milky Way, $\Delta t \sim 1$ 
Gyr, which is comparable to the observed age spread.  
Additionally,  Freeman \& Bland-Hawthorn (2002)    have remarked on 
the fact that halo GCs 
and stars in the thick disk of our Galaxy have very similar abundances and
ages.  While some of these disk stars may be from globulars long
since destroyed, according to our picture, others may have
formed from gas in outflow-enriched protocluster clouds 
for which the conditions were never conducive to GC formation.

The small range in metallicities amongst the stars {\em within} any
given cluster points to an external mechanism for enrichment, as
provided in our picture.   Furthermore the time scale for this
enrichment must be short, as the presence of dust and metals will
inevitably spark cooling and star formation, which will further enrich
material that is not quickly formed into stars.   Outflow-minihalo
interactions have the  potential to achieve such rapid mixing,
but order of magnitude estimates can not confirm this definitively.
Rather the problem of mixing is a delicate one, in which the time
scale for star formation in shocked regions is comparable to the
global dynamical time, and the detailed turbulent structure is key.
In the numerical studies presented in this paper we have taken a first
step towards improving these estimates,  exploring a simple model of
turbulent mixing in  a one-dimensional context. Yet the 
$\xi \gtrsim 20\%$  shear-induced mixing seen in this study 
highlights the importance of large scale vortices in turbulence
generation.  Multi-dimensional simulations are necessary to move
further, and a definitive answer will require detailed modeling of
cooling, gravitation, and perhaps even the turbulent spectral
distribution.

Although the issue of mixing is complicated, two properties of the
metals available in  outflow-minihalo interactions are
suggestive of recent GC observations. First, as pointed out by
Beasley \etal (2003), no single Pop III object seems capable of
producing the typical abundance ratios of heavy elements in GC stars.
Some distribution of first generation stars is needed, and a galactic
outflow is the likely consequence of several concurrent supernovae.
Second, the existence of a correlation between a parent galaxy's
luminosity and the mean color (viz., metallicity) of its GC systems
(Strader, Brodie, \& Forbes 2004) argues strongly against
self-enrichment as the source of a GC's heavy elements.

Our basic model even enables us to deduce such a
dependence.  As the efficiency of our scenario is primarily
due to an interplay between cooling and sound-crossing times,
both which depend on the shock velocity, we fix a typical $v_s$ 
at which GC formation is effective.
From eq.\ (\ref{eq:vs}) this gives that $R_s \propto
E_{55}^{1/3} (1+z_s)^{-1}.$  We then assume that, on average,  the
total mass in metals and the outflow energy 
will be  proportional to luminosity of the parent galaxy to which the
resulting globular cluster belongs, yielding  
$Z_c  \propto L_{\rm gal}^{1/3}
(1+z_s)^2.$  From the simple top-hat collapse model
$(1+z_s)  \propto (M_{\rm gal}) ^{-(n+3)/6}$, where
$n$ is the slope of the primordial power spectrum, so again taking the
mass proportional to $L_{\rm gal}$, we obtain the prediction
\be
Z_c  \propto L_{\rm gal}^{-(n+2)/3},
\ee
where for the CDM model $n$ is between -3 and and -2 for the scales 
of interest.
This gives a final index ranging from 1/3 for small galaxies
to 0 for the largest ones, bracketing the observed value of 
1/6 (Strader, Brodie, \& Forbes 2004).

A similar calculation can be carried out to estimate the maximum 
galactocentric radius of the distribution of halo GCs.  
Fixing $v_s$ we have 
\be
R_{s,{\rm max}} \propto E^{1/3} (1+z_s)^{-1} \propto L_{\rm gal}^{(n+5)/6}
\ee
such that for the $n$ values appropriate for CDM,  the index
ranges from ${1/3}$ for small galaxies to ${1/2}$  for large galaxies,
and the ``best-fit'' $n$ value to the observed  metallicity-luminosity
relationship gives $R_{s,{\rm max}} \propto L^{5/12}$.  As such
observations require both high-angular resolution and a wide field of
view,  no current constraints on this relationship  are available in
the literature.  However, such measurements are well within the
capabilities of the {\em Advanced Camera for Surveys} on the {\em Hubble
Space Telescope} and new information in this area should be  expected
to become available soon.  Note that this scaling with luminosity is likely
to persist even if substantial ``shuffling'' of globular clusters {\em
within} this maximum radius is likely to have taken place, as a large
fraction of halo GCs are on close to radial orbits (eg.\ Dinescu \etal
1999;  Wang \& Zhou 2003).  
On the other hand, any initial metallicity gradient established
among a galaxy's halo GCs is expected to be washed-out by this
shuffling and possibly by the addition of clusters striped
from small, neighboring galaxies.   It is also
interesting to note that while no clear metallicity
gradients have ever been observed in halo GCs, there are hints of a
fall off in metallicity with radius in the Milky Way Halo GC
population when one attempts to remove accreted systems
(Zinn 1993; Parmentier \etal 2001).

While these comparisons are perhaps a bit simplified, it is
reassuring that recent investigations not only uncovered a trend with
luminosity, but one compatible with our model, and that we are able to
construct a simple prediction that will be testable in the near
future.  Perhaps even more reassuring is the fact that the ideas
presented here allow for immediate points of contact between theory
and new observations.  Although many issues remain to be better
understood, our scenario provides clear directions that can lead to
more detailed comparisons with various observational constraints.
Such future studies may soon shed new light on the old question of the
origin of halo globular clusters.

\acknowledgements 

\begin{figure}
\centerline{\psfig{figure=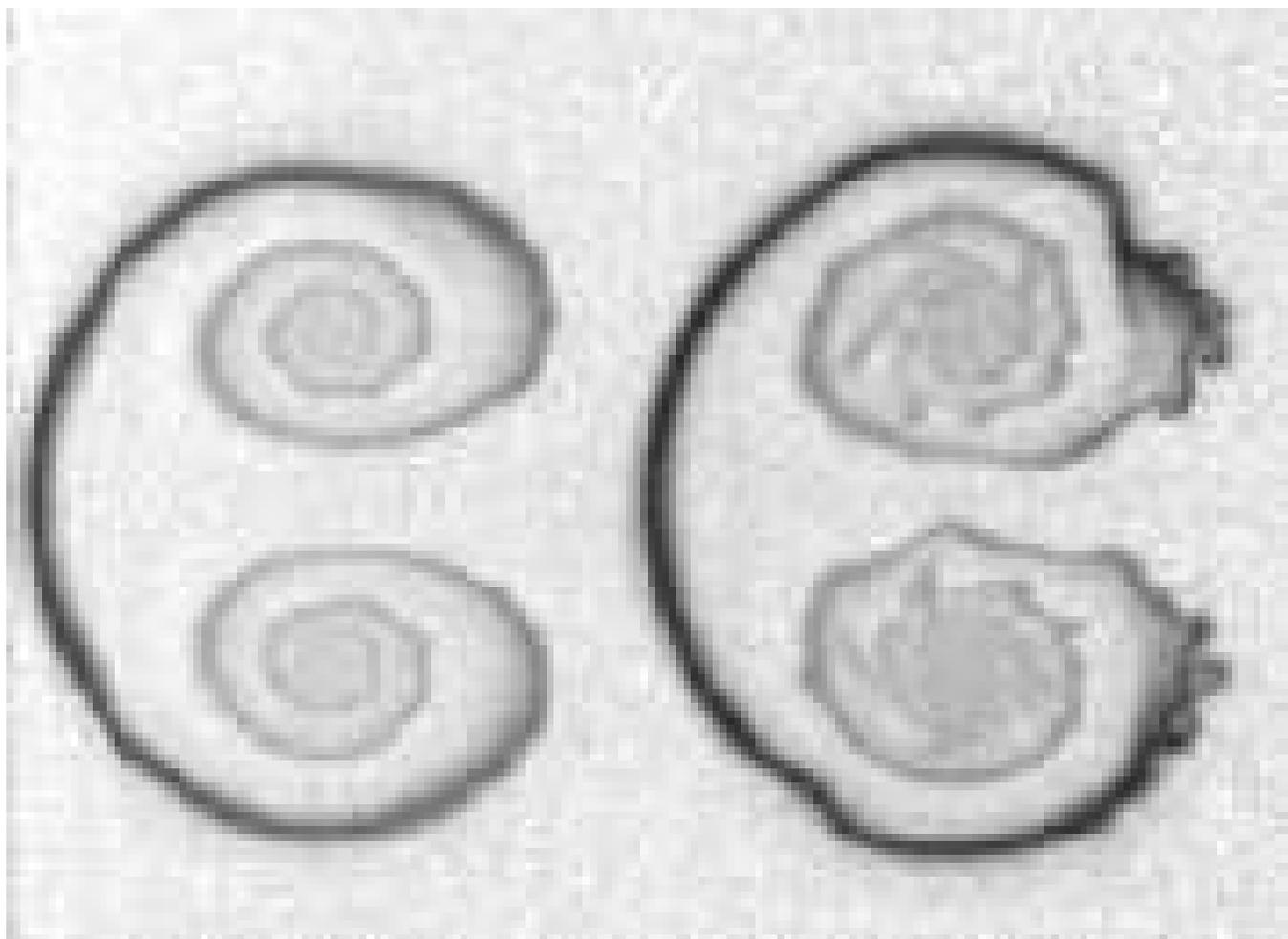,height=13cm}}
\caption{Experimental results of a column of heavy gas ($SF_6$)
which is overtaken by a planar shock, which moves in from the 
left in these figures.  The column is 3.1 mm in diameter and
time frames are at 320 (left) and 470 (right) microseconds.
In our units of radius over the shock speed, these correspond 
to times of 90 and 130 repectively.
This image was provided by C. D.\ Tomkins and was taken
from the experiments described in Vorobieff \etal (2003).}
\label{fig:experiment}
\end{figure}

We are grateful to Jean Brodie, Alex Heger, Chris McKee, Francesco
Palla, Jay Strader, and an anonymous referee
for helpful comments during the preparation of 
this manuscript.
Thanks with much appreciation to Christopher D.\ Tomkins,
Hydrodynamics Group DX-3, Los Alamos National Laboratory, who provided
us with the experimental results presented in Figure
\ref{fig:experiment}.  ES was supported by an NSF Math and Physical
Sciences Distinguished International Postdoctoral Research (NFS
MPS-DRF)  fellowship during part of this investigation; his research
was also supported by the National Science Foundation under grant
PHY99-07949.  The work performed by JW and FH was conducted under
the auspices of the U. S. Department of Energy's contract
W-7405-ENG-36 with the University of California.

\fontsize{10}{10pt}\selectfont

\newpage

\end{document}